\documentclass[12pt,letterpaper]{article} 
\usepackage{jheppub}

\usepackage{graphicx} 
\usepackage{caption}
\usepackage{subcaption}

\graphicspath{{../../Figures/}}

\usepackage{amsmath}

\title{A Quantum Focussing Conjecture}

 \author[a,b]{Raphael Bousso,}
\author[a,b]{Zachary Fisher,} 
\author[a,b]{Stefan Leichenauer,} 
 \author[c]{and Aron C. Wall} 
\affiliation[a]{Center for Theoretical Physics and Department of Physics,\\
University of California, Berkeley, CA 94720, U.S.A.} 
\affiliation[b]{Lawrence Berkeley National Laboratory, Berkeley, CA 94720, U.S.A.} 
 \affiliation[c]{Institute for Advanced Study, Princeton, NJ 08540, USA}

\abstract{We propose a universal inequality that unifies the Bousso bound with the classical focussing theorem. Given a surface $\sigma$ that need not lie on a horizon, we define a finite {\em generalized entropy} $S_\text{gen}$ as the area of $\sigma$ in Planck units, plus the von Neumann entropy of its exterior. Given a null congruence 
$N$ 
orthogonal to $\sigma$, the rate of change of $S_\text{gen}$ per unit area defines a {\em quantum expansion}. We conjecture that the quantum expansion cannot increase 
along $N$. 
This extends the notion of universal focussing to cases where quantum matter may violate the null energy condition. Integrating the conjecture yields a precise version of the Strominger-Thompson {\em Quantum Bousso Bound}. Applied to locally parallel light-rays, the conjecture implies a {\em Quantum Null Energy Condition}: a lower bound on the stress tensor in terms of the second derivative of the von Neumann entropy. We sketch a proof of this novel relation in quantum field theory.}

\begin{document} 
\maketitle 

\section{Introduction}
\label{sec-intro}

The study of black holes has revealed a profound connection between quantum information and geometry, beginning with Bekenstein's proposal~\cite{Bek72,Bek73,Bek74} that black holes have an entropy proportional to their horizon area:
\begin{equation}
S_{\rm BH}=\frac{A}{4G\hbar}~.
\end{equation}
The coefficient $1/4$ is fixed by the First Law of thermodynamics, $dE = T dS$, using Hawking's later calculation of the black hole temperature~\cite{Haw74a,Haw75}.

Moreover, Bekenstein introduced the notion of {\em generalized entropy}, defined as the sum of black hole horizon entropy and the ``ordinary'' entropy of matter systems in the exterior of black holes, $S_\text{out}$:
\begin{equation}
S_\text{gen} = \frac{A}{4G\hbar} + S_\text{out} ~.
\label{eq-sgenintro}
\end{equation}
Using this definition, Bekenstein proposed a {\em Generalized Second Law} of thermodynamics (GSL), which states that the generalized entropy will not decrease in any physical process:
\begin{equation}
d S_\text{gen} \geq 0~.
\end{equation}
The GSL was introduced as a successor to the ordinary Second Law, which fails when matter entropy disappears into a black hole. The GSL proved to be a successor to the classical area law for black hole horizons~\cite{Haw71}, as well: when a black hole evaporates, its area decreases; but the Hawking radiation produced in the exterior more than compensates~\cite{Pag76}, so the GSL is upheld.

The connection between quantum information and geometry was deepened by the recognition that the GSL appears to impose limitations on the entropy content of weakly gravitating matter systems~\cite{Bek81} and of certain spacetime regions. This led to the proposal of a {\em holographic principle}~\cite{Tho93,Sus95,FisSus98,CEB1,CEB2,RMP}, which holds that the information (or entropy) content of spacetime regions is fundamentally governed by the area of surfaces, and that this apparently nonlocal constraint will be manifest in a full quantum gravity theory. The AdS/CFT correspondence~\cite{Mal97} partly fulfills this expectation~\cite{SusWit98} in a certain class of spacetimes.


The covariant entropy bound (or Bousso bound) relates matter entropy to the area of arbitrary surfaces, not just black hole horizons~\cite{CEB1}. In the generalized form of~\cite{FMW} the bound states that 
\begin{equation}
\Delta S \leq \frac{\Delta A}{4G\hbar}~,
\label{eq-cebintro}
\end{equation}
where $\Delta S$ is the matter entropy passing through a nonexpanding null hypersurface bounded by surfaces whose areas differ by $\Delta A$. More details are given in Sec.~\ref{sec-classical}; for a full review, see~\cite{RMP}.

The Bousso bound reduces to previous heuristic bounds in well-circumscribed settings~\cite{CEB1,FMW,Bou03,BCFM1}. But its validity extends to cosmological spacetimes and regions deep inside a black hole. Since such regions cannot be converted to black holes of the same area, the bound hints at a much broader relation between quantum information and geometry that goes beyond black hole thermodynamics. 

In a similar vein, we will argue here that a generalized entropy should be ascribed not only to black hole and other causal horizons \cite{JP03}, but to a much larger class of surfaces~\cite{Wall10,BianchiMyers12,MyePou13,FLM13,EngWal14}. Clearly, Eq.~(\ref{eq-sgenintro}) allows us to assign a generalized entropy to any surface $\sigma$ that divides a Cauchy surface into two portions, where $S_\text{out}$ can be taken to be the matter entropy on either one of these portions. We will find that this viewpoint leads to a statement more powerful than the Bousso bound: the Quantum Focussing Conjecture.

The generalized entropy is a promising and versatile notion because it is finite. Newton's constant $G$ and the exterior entropy are separately cutoff-dependent. But over the past decades, evidence has mounted\footnote{For example, the leading divergence of the vacuum entanglement entropy near the surface $\sigma$ scales like $A/\epsilon^2$ but $1/G$ is renormalized so as to absorb this divergence. Subleading divergences of the entanglement are similarly cancelled by appropriate higher-curvature corrections to the gravitational action. We give a review of these results along with extensive references in Appendix~\ref{ren}.} that the combined divergences cancel, leaving a finite piece that is invariant under RG flow, as originally proposed in \cite{Susskind:1994sm} and expanded upon in \cite{Jacobson:1994iw, Frolov:1996aj}).  What we call gravitational entropy and matter entropy depends on the cutoff; but their sum, $S_\text{gen}$, does not. This suggests that $S_\text{gen}$, unlike $S_\text{out}$ or $A/4G\hbar$ separately, reflects some information present in the full quantum gravity theory. Presumably, $S_\text{gen}$ is a measure of the entropy of the degrees of freedom accessible on one side of that surface, where the area term represents the dominant contribution coming from Planckian degrees of freedom very close to $\sigma$, somehow cut off by quantum gravity \cite{Sorkin83, Frolov:1993ym, Barvinsky:1994jca, Susskind:1994sm, Jacobson:1994iw, Frolov:1996aj}.

However, the generalized entropy is only a semiclassical concept, assigning to each surface an entropy proportional to its area, without worrying about whether these degrees of freedom are entangled with each other, or with other systems.  Thus it does not capture nonperturbative physics such as the (presumed) unitarity of Hawking evaporation. For example, the GSL does not hold (except in a coarse-grained sense) after the Page time \cite{Page:1993wv}, when more than half the entropy has radiated out of the black hole, so that the hidden purity becomes potentially measurable.  At this stage, both the fine-grained entropy of the exterior and the area of the horizon decrease, and the correct statement of the second law becomes different.  Similarly, the Quantum Focussing Conjecture is a semiclassical statement that may need to be modified in the nonperturbative regime, e.g., for sufficiently old black holes that have information on the horizon.  In this article we will confine our analysis to situations where the semiclassical analysis is valid.

At a practical level, extending the notion of generalized entropy to arbitrary surfaces yields powerful extensions of classical GR results to the semiclassical level (much as the GSL supersedes the classical area theorem for causal horizons). For example, Penrose's singularity theorem for trapped surfaces~\cite{Pen65} fails for evaporating black holes because it cannot accommodate quantum fluctuations with negative energy. But a more robust theorem guarantees singularities in the presence of quantum trapped surfaces, which are defined in terms of the generalized entropy~\cite{Wall10}.

Here we will use $S_\text{gen}$ to formulate an extension of the classical focussing theorem for surface-orthogonal null congruences. The classical theorem states that light-rays never ``anti-focus'' as long as matter has positive energy. Mathematically, the expansion scalar $\theta$ cannot increase along a congruence of lightrays, where $\theta$ is the logarithmic derivative of the area spanned by the light-rays:
\begin{equation}
\frac{d\theta}{d\lambda} \equiv \frac{d}{d\lambda}\left(\frac{d{\cal A}/d\lambda}{\cal A}\right) \leq 0~,
\label{eq-clfocintro}
\end{equation}
where ${\cal A}$ is an infinitesimal area element spanned by nearby null geodesics, and $\lambda$ is an affine parameter. We review this result in Sec.~\ref{sec-classical}.

Because quantum fluctuations can have negative energy, the classical focussing theorem fails at the semiclassical level (e.g., near black hole horizons), just as the area theorem and the singularity theorem fail. In Sec.~\ref{sec-qefc}, we define a {\em quantum expansion}, $\Theta$, as a functional derivative (per unit area) of the generalized entropy along a null congruence orthogonal to the surface $\sigma$. We conjecture that $\Theta$ cannot increase along any congruence, even in quantum states that would violate the classical focussing theorem:
\begin{equation}
\frac{d\Theta}{d\lambda} \leq 0~.
\end{equation}
This is the Quantum Focussing Conjecture (QFC).

We derive and explore two important implications of the QFC. In Sec.~\ref{sec-qceb}, we show that the QFC implies the Bousso bound, but in an improved form. The Bousso bound was initially formulated only for the case where the matter entropy is dominated by isolated systems. In this setting, a finite entropy is easily computed from a density operator for the system, or by integrating an entropy density. In more general settings, the matter entropy cannot be cleanly separated from the divergent vacuum entanglement entropy across the surface $\sigma$. Two inequivalent quantum extensions of the bound have been put forward: in the weakly gravitating regime, the entropy can be regulated by vacuum subtraction~\cite{HolLar94,MarMin04,Cas08,BCFM1,BCFM2}; in a more general setting, one must include the vacuum entanglement~\cite{StrTho03}. 

We recover a ``quantum Bousso bound'' by integrating the QFC. On a (quantum) light-sheet, the generalized entropy is initially decreasing, so by the integrated QFC it will continue to decrease. Hence, the initial generalized entropy is greater than the final one. This statement is manifestly cutoff-independent, i.e., it is automatically equipped to deal with the divergences of the von Neumann entropy. It is closely related to an early improvement of the Bousso bound by Strominger and Thompson~\cite{StrTho03}. Breaking $S_\text{gen}$ into $S_\text{out}+A/4G\hbar$ and rearranging terms, one recovers the Bousso bound in the familiar form of Eq.~(\ref{eq-cebintro}), $\Delta S\leq \Delta A/4G\hbar$; see Fig.~\ref{fig-nowife}.

In Sec.~\ref{sec-qnec} we explore the QFC in settings where the classical expansion vanishes. We find that the QFC implies a novel {\em Quantum Null Energy Condition}, 
\begin{equation}
T_{kk}\geq \lim_{{\cal A}\to 0}~ \frac{\hbar}{2\pi {\cal A}}~ \frac{d^2 S_\text{out}}{d\lambda^2}~,
\label{eq-qnecintro}
\end{equation}
where $T_{kk}$ is the null-null component of the stress tensor. We sketch a proof of the Quantum Null Energy Condition for free fields in Minkowski space. This provides significant evidence supporting the QFC, beyond the evidence already supporting the Bousso bound~\cite{CEB1,FMW,RMP,BouFla03,StrTho03,BCFM1,BCFM2}.

In Sec.~\ref{sec-other}, we discuss how the QFC relates to other proposals and results, including the GSL, the quantum Bousso bounds of~\cite{StrTho03} and of~\cite{BCFM1,BCFM2}, the quantum singularity theorem of~\cite{Wall10}, the quantum extremal surface barriers of~\cite{EngWal14}, and a novel GSL for quantum holographic screens~\cite{BouEng15a,BouEng15b,BouEngTA}.

\section{Classical Focussing and Bousso Bound}
\label{sec-classical} 

In this section we review the classical notion of the expansion of a null congruence, and two statements that involve this expansion: the classical focussing theorem, and the Bousso bound. Both will later be subsumed by the Quantum Focussing Conjecture.

\subsection{Classical Expansion}
\label{sec-classexp}

Consider a congruence of light rays emanating orthogonally from a codimension-2 spacelike hypersurface. The {\em expansion scalar} $\theta$ is defined as the trace of the null extrinsic curvature~\cite{Wald}
\begin{equation}
\theta\equiv \nabla_a k^a~.
\end{equation}
Here $k^a=(d/d\lambda)^a$ is the (null) tangent vector to the congruence, normalized with respect to an affine parameter $\lambda$. Equivalently, $\theta$ is the logarithmic derivative of the area element $\mathcal{A}$ spanned by infinitesimally neighboring geodesics: 
\begin{equation}
  \label{eq-expand}
\theta = \lim_{\mathcal A \rightarrow 0} \frac{1}{\cal A} \frac{d{\cal A}}{d\lambda}~.
\end{equation}
From its definition, it is clear that $\theta$ is a local quantity. 

A {\em caustic (conjugate point, focal point)} is a point where $\theta\to -\infty$, which happens when the cross-sectional area vanishes, i.e., when infinitesimally neighboring geodesics intersect.

The evolution of the expansion $\theta$ along the congruence is determined by the Raychaudhuri equation:
\begin{equation}
  \frac{d\theta}{d\lambda} = -\frac{1}{D-2}\theta^2 - \sigma_{ab}\sigma^{ab} - R_{ab} k^a k^b~,
\label{eq-raych}
\end{equation}
where $R_{ab}$ is the Ricci tensor and $D$ is the spacetime dimension. The shear $\sigma_{ab}$ is defined as the trace-free symmetric part of the null extrinsic curvature~\cite{Wald}.

\subsection{Classical Focussing Theorem}\label{sec-classicalfocus}
\label{sec-classfocus}

In a spacetime which satisfies the null curvature condition, namely $R_{ab} k^a k^b \ge 0$ for all null vectors $k^a$,  each term on the right-hand side of \eqref{eq-raych} is manifestly nonpositive. Physically, this means that light rays can focus but not anti-focus, and it implies the following theorem:\footnote{There exists another version of the focussing theorem, which also follows from Eq.~(\ref{eq-raych}) and the null curvature condition: if $\theta(p)$ is strictly negative at some point $p$ on a null geodesic, then there will be a caustic on the geodesic, at affine parameter no further than $|(D-2)/\theta(p)|$ from the point $p$. We will not consider this theorem here, because its quantum generalization is not yet known.}

{\em In a spacetime satisfying the null curvature condition, the expansion is nonincreasing at all regular (non-caustic) points of a surface-orthogonal null congruence:}
\begin{align}
  \frac{d\theta}{d\lambda} \le 0~.
\label{eq-clfoc}
\end{align}


In Einstein gravity, the null curvature condition is equivalent to the null energy condition, that the stress tensor obeys $\langle T_{ab} \rangle k^a k^b\geq 0$ for all null vectors $k^a$. This inequality is broadly obeyed in the classical limit, and by coherent states of a quantum field. 

However, the null energy condition is not universally valid. It is violated by physically reasonable states in the quantum field theory \cite{EGJ65}, for example by the Casimir effect \cite{Casimir:1948dh}, moving mirrors \cite{Davies:1976hi, Davies:1977yv}, squeezed states of light \cite{Braunstein, MT88}, and Hawking radiation \cite{Haw75, Davies:1976ei}.  In any region where it is violated, one can construct a counterexample to the above focussing theorem, by choosing a congruence with sufficiently small $\theta$ and $\sigma_{ab}$.

\subsection{Bousso Bound}
\label{sec-classceb}

A light-sheet is a null hypersurface with everywhere nonpositive expansion $\theta\leq 0$. The Bousso bound~\cite{CEB1,CEB2} is the conjecture that the entropy on a light-sheet cannot exceed the area of its initial cross-section in Planck units, $A/4G\hbar$. The bound can be strengthened~\cite{FMW} in the case where the light-sheet is truncated at some nonzero final area $A'$:
\begin{equation}
S\leq \frac{A-A'}{4G\hbar}~,
\label{eq-gceb}
\end{equation}

The Bousso bound is useful in regimes where the quantities it relates are well-defined. In the semiclassical regime, the areas of surfaces are sharply defined. In many situations, the entropy is also easy to compute, for example when dealing with well-isolated matter systems, or with a portion of an extensive system large enough for the notion of entropy density to be meaningful. 

In the semiclassical regime, the areas of surfaces are sharply defined. In many situations the entropy is also easy to compute, for example when dealing with well-isolated matter systems, or with a portion of an extensive system large enough for the notion of entropy density to be meaningful. Within this wide arena, there exist strong counterexamples to all alternative proposals (so far) of the general form $S\lesssim A/4G\hbar$~\cite{KalLin98}. The Bousso bound evades these counterexamples because of the special properties of light-sheets. Thus, the notion of light-sheets (rather than spatial volumes, or light-cones lacking a nonexpansion condition~\cite{FisSus98}), appears to be crucial.

It would be nice to broaden the regime for which the entropy $S$ is well-defined. A clarification is particularly necessary in the case where the Bousso bound is applied to a system consisting of only a few quanta, such as a single photon wavepacket with Gaussian profile. Globally, the entropy will be of order unity (assuming, for example, an incoherent superposition of different polarization states).  However, it is not obvious how to define the entropy on a finite light-sheet: some tail of the wavepacket will be missing, so one cannot use the global density matrix. Restricting the density operator to the finite light-sheet, the von Neumann entropy receives a divergent contribution from entanglement across the boundaries at the two surfaces $A$ and $A'$. This contribution dominates but it is intuitively unrelated to the photon.

A sharp definition of $S$ was given recently for light-sheets in the weak gravity limit, $G\hbar\to 0$,
with perturbative matter.
In this regime one can restrict both the vacuum and the state of interest to the same region or light-sheet. In this setting, the entropy $S$ can be defined as the difference of the two resulting von Neumann entropies of the light-sheet states~\cite{HolLar94,MarMin04,Cas08}. Because the divergences of the entanglement entropy are associated with its boundary, this quantity is finite and reduces to the expected entropy for isolated systems and fluids.\footnote{In the interacting case, it reduces to an upper bound on the naive entropy, which suffices.} With this definition, the bound can be proven~\cite{BCFM1,BCFM2} to hold in the weak gravity limit.

However, this definition cannot be applied when gravity is strong, since it is not clear what one would mean by the ``same'' light-sheet for two states with different geometry. It is therefore necessary to find some other definition of entropy such that a Bousso bound can be precisely formulated (and perhaps proven).


We will show below that the \emph{Quantum Focussing Conjecture} furnishes such a definition. Interestingly, we will find that this definition does not reduce to that of~\cite{BCFM1,BCFM2} in the weak-gravity limit, where the latter is well defined.  Moreover, we will find that the nonexpansion condition $\theta\leq 0$, which was strictly preserved in~\cite{BCFM1,BCFM2}, will be modified to a ``quantum nonexpansion condition''.  The resulting conjecture is similar to that of \cite{StrTho03}; we will comment on the differences in Sec.~\ref{sec-ST}.

\section{Quantum Expansion and Focussing Conjecture}
\label{sec-qefc}

In this section, we define the notion of {\em quantum expansion} as a functional derivative of the generalized entropy, and we formulate the Quantum Focussing Conjecture.

\subsection{Generalized Entropy for Cauchy-splitting Surfaces}
\label{sec-genen}

Generalized entropy was originally defined~\cite{Bek72} in asymptotically flat space, as the area $A$ of all black hole horizons (in Planck units), plus the entropy of matter systems outside the black holes:
\begin{equation}
S_\text{gen} \equiv S_\text{out} + \frac{A}{4G\hbar} +\mathrm{counterterms}~.
\label{eq-sgendef}
\end{equation}
A rigorous definition of $S_\text{out}$ can be given as the von Neumann entropy of the quantum state of the exterior of the horizon: 
\begin{equation}
S_\text{out} = - \mathrm{tr}\, \rho_{\rm out} \log \rho_{\rm out}~.
\label{eq-soutdef}
\end{equation}
The reduced density matrix $\rho_{\rm out}$ is obtained from the global quantum state $\rho$ by tracing out the field degrees of freedom behind the horizon:
\begin{equation}
\rho_{\rm out} = \mathrm{tr}_{\rm in}\, \rho~.
\label{eq-routdef}
\end{equation}
(If the global state of the matter fields is pure, then $S_\text{out} =S_\text{in}$, where $S_\text{in}$ is the von Neumann entropy of the interior region.)

The von Neumann entropy $S_\text{out}$ is UV-divergent. However, there is now strong evidence that this divergence is precisely cancelled by a renormalization of Newton's constant in the area term. Subleading divergences are cancelled by other geometric counterterms. This is discussed in detail in Appendix~\ref{ren}; here we assume that the generalized entropy is indeed finite and cutoff-independent. 

The generalized entropy was introduced in order to salvage the second law of thermodynamics when matter entropy is lost into a black hole. Bekenstein conjectured that a Generalized Second Law~\cite{Bek72} (GSL) survives: the area increase of the black hole horizon will compensate or overcompensate for the lost matter entropy, so that the generalized entropy will not decrease. The GSL does appear to hold for realistic matter entering a black hole, and it has been proven in certain settings (for recent proofs see \cite{Wall11, Sarkar:2013swa, Bhattacharjee:2015yaa, Wall:2015raa}, and for a review of previous proofs, see \cite{Wall09} and references therein).

The GSL supersedes not only the ordinary second law, but also Hawking's area theorem. When a black hole evaporates~\cite{Haw75}, the null energy condition is violated, and the area decreases. However, the emitted radiation more than compensates for this decrease~\cite{Pag76}. 

We now follow Refs.~\cite{Wall10,MyePou13,EngWal14} and extend the notion of generalized entropy beyond the context of causal horizons. Let $\sigma$ be a spacelike codimension-2 surface $\sigma$ that splits a Cauchy surface $\Sigma$ into two portions. The surface $\sigma$ need not be connected; for example, it may be the union of several black hole horizons, or of two concentric topological spheres. Nor does it need to be compact; for example, it could be a cross-section of a Rindler horizon. 

\begin{figure}
	\centering
    \begin{subfigure}[b]{2.4in}
            \includegraphics[width=2.2in]{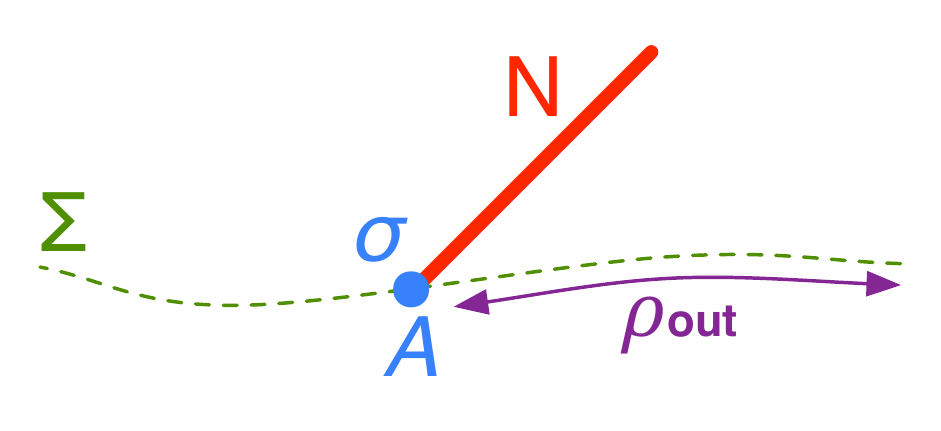}
            \caption{}
            \label{fig-cauchy1}
    \end{subfigure}
	\hspace{.5in}
       \begin{subfigure}[b]{1.6in}
            \includegraphics[width=1.6in]{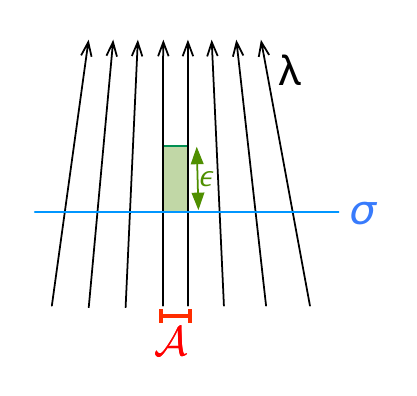}
            \caption{}
            \label{fig-cauchy2}
    \end{subfigure}
	\caption{(a) A spatial surface $\sigma$ of area $A$ splits a Cauchy surface $\Sigma$ into two parts. The generalized entropy is defined by $S_\text{gen} = S_\text{out}+ A/4G\hbar$, where $S_\text{out}$ is the von Neumann entropy of the quantum state on one side of $\sigma$. To define the quantum expansion $\Theta$ at $\sigma$, we erect an orthogonal null hypersurface $N$, and we consider the response of $S_\text{gen}$ to deformations of $\sigma$ along $N$. (b) More precisely, $N$ can be divided into pencils of width $\mathcal A$ around its null generators; the surface $\sigma$ is deformed an affine parameter length $\epsilon$ along one of the generators, shown in green.}
	\label{fig-thetadef}
\end{figure}

We pick one of the two sides of $\sigma$ arbitrarily and refer to it as $\Sigma_\text{out}$; see Fig.~\ref{fig-cauchy1}. We use Eqs.~(\ref{eq-sgendef}-\ref{eq-routdef}) to define a generalized entropy.

This viewpoint has two important consequences. Suppose we are given any theorem or conjecture about the area of surfaces (such as Hawking's area theorem), which is valid classically, but which can fail when the null curvature condition is violated. By a judicious application of the substitution
\begin{equation}
A \to 4G\hbar\, S_\text{gen} ~,
\label{eq-sub}
\end{equation}
we may obtain a semi-classical statement of much broader validity (such as the GSL). Indeed, the notion of generalized entropy of non-horizon surface has been profitably applied to Penrose's singularity theorem~\cite{Wall10}, to the Ryu-Takayanagi proposal~\cite{FLM13,EngWal14}, and to a novel area law for holographic screens~\cite{BouEng15a,BouEng15b,BouEngTA}. Below, we will apply Eq.~(\ref{eq-sub}) to the classical focussing theorem, Eq.~(\ref{eq-clfoc}), to obtain a Quantum Focussing Conjecture.

Secondly, the quantity $S_\text{gen}$ provides a cutoff-independent measure of entropy in a bounded region, because the geometric terms cancel the divergences of the von Neumann entropy. Unlike vacuum subtraction, this feature does not rely on a weak-gravity limit.  We will exploit this to formulate a quantum Bousso bound in terms of the generalized entropy. In fact, we find that an appropriate bound arises simply as a special case of our Quantum Focussing Conjecture, which we will now formulate.

\subsection{Quantum Expansion}
\label{sec-qexp}

As a first step, let us use the substitution (\ref{eq-sub}) to define a {\em quantum expansion}. We now need an additional structure: the null geodesics orthogonal to $\sigma$ that define the classical expansion. In addition to the twofold choice of $\Sigma_\text{out}$, this faces us with an additional, fourfold choice: there are four null hypersurfaces orthogonal to $\sigma$. They are generated by orthogonal light-rays towards the past or future and towards $\Sigma_L$ or $\Sigma_R$ (regardless of which one is chosen as $\Sigma_\text{out}$). Again, we may pick any direction, e.g., the one shown in Fig.~\ref{fig-cauchy1}. (All of the statements below will hold under any of the eight possible choices; in particular, the QFC will be conjectured to hold at this broad level.)

The chosen hypersurface $N$ will be terminated by caustics, or more generally wherever null generators orthogonal to $\sigma$ intersect. Thus, $N$ consists of one component of the boundary of the past, or of the future, of $\sigma$~\cite{Wald}.

Through each point $y$ of $\sigma$ there passes one generator of $N$; see Fig.~\ref{fig-cauchy2}. We take $\lambda$ to be an affine parameter along this generator, such that $\lambda=0$ on $\sigma$ and $\lambda$ increases away from $\sigma$. This defines a coordinate system ($\lambda,y$) on $N$. 

A positive definite function $V(y)\geq 0$ defines a slice of $N$, consisting of the point on each generator $y$ for which $\lambda=V$. Any such slice of $N$ splits a Cauchy surface into two parts. Hence $V(y)$ is the argument of a generalized entropy functional
\begin{equation}
S_\text{gen}[V(y)] = \frac{A[V(y)]}{4G\hbar} + S_\text{out}[V(y)]~.
\end{equation}

The quantum expansion, like the classical expansion, is defined by deforming a slice in the neighborhood of one generator $y_1$. To be precise, consider a second slice of $N$ which differs from $\sigma$ only in a neighborhood of generators near $y_1$, with infinitesimal area ${\cal A}$:
\begin{equation}
V_\epsilon(y) \equiv V(y) + \epsilon\,\vartheta_{y_1}(y)~.
\end{equation} 
Here we define $\vartheta_{y_1} = 1$ in a neighborhood of area ${\cal A}$ around a point $y_1$, and $\vartheta_{y_1}=0$ everywhere else; see Fig.~\ref{fig-cauchy2}.  One can differentiate the generalized entropy with respect to this localized deformation:
\begin{equation}
\left.\frac{dS_\text{gen}}{d\epsilon}\right|_{y_1} \equiv \lim_{\epsilon\to 0} \frac{S_\text{gen}[V_\epsilon(y)] - S_\text{gen}[V(y)]}{\epsilon}~.
\end{equation}
The quantum expansion is the finite quantity obtained by dividing this derivative by the infinitesimal unit area ${\cal A}$, just as in the classical case \eqref{eq-expand}:
\begin{equation}\label{eq-thetadiscrete}
\Theta[V(y);y_1] \equiv \lim_{{\cal A} \to 0} \left.\frac{4G\hbar}{\cal A} \,
\frac{dS_\text{gen}}{d\epsilon}\right|_{y_1}~.
\end{equation}

The above construction is equivalent to defining $\Theta$ as the functional derivative of $S_\text{gen}$ with respect to $V(y)$:
\begin{equation}
\Theta[V(y);y_1] \equiv \frac{4G\hbar}{\sqrt{^V\!g(y_1)}} \frac{\delta S_\text{gen}}{\delta V(y_1)}~,
\end{equation}
where $\sqrt{^V\!g}$ is the (finite) area element of the metric restricted to $\sigma$, inserted to ensure that the functional derivative is taken per unit geometrical area, not coordinate area. The notation $\Theta[V(y);y_1]$ emphasizes that the quantum expansion requires the specification of a slice $V(y)$ and is a function of the coordinate $y_1$ on that slice.

The classical expansion $\theta$ depends only on the infinitesimal neighborhood of a null generator. By contrast, the quantum expansion $\Theta$ depends nonlocally on the quantum state of matter on the half-Cauchy-surface $\Sigma_\mathrm{out}$, because the von Neumann entropy of the matter can behave differently at $y_1$ if one changes the state of matter elsewhere on $\Sigma_\text{out}$. Moreover, the quantum expansion at $y_1$, $\Theta[V(y),y_1]$, depends on the choice of $V(y)$ away from $y_1$.  However, $\Theta$ does not depend on the choice of the half-Cauchy-surface attached to the spatial slice $V(y)$ of $N$, since all Cauchy surfaces are unitarily equivalent. This freedom makes it possible to find a suitable $\Sigma_\text{out}$ for any deformation $V(y)$; note that portions of $\Sigma_\text{out}$ may coincide with $N$ without violating the achronality condition on Cauchy surfaces.

In the classical limit $G\hbar \to 0$
with the classical geometry held fixed,
the matter entropy $S_\text{out}$ does not contribute, and $\Theta$ reduces to the local geometric expansion $\theta$ at each generator $y_1$.

\subsection{Quantum Focussing Conjecture}

The definition of a quantum expansion allows us to formulate a generalization of the classical focussing theorem, Eq.~(\ref{eq-clfoc}), to the semiclassical regime. The quantum focussing conjecture (QFC) is the statement that
\begin{equation}
\frac{\delta}{\delta V(y_2)} \Theta[V(y); y_1] \le 0~.
\label{eq-sqfc}
\end{equation}
In words, {\em the quantum expansion cannot increase at $y_1$, if the slice of $N$ defined by $V(y)$ is infinitesimally deformed along the generator $y_2$ of $N$}, in the same direction. Here $y_2$ can be taken to be either the same or different from $y_1$. 

The QFC is nonlocal: as noted above, the generalized entropy depends on all of $\sigma$, and so do its first and second functional derivatives. Even the sign of $\Theta[V(y)]$ at some point may depend on the choice of $V$ away from this point. 

We defined the QFC so that it applies regardless of which side of $N$ we choose to compute the generalized entropy and its derivatives. In principle one could distinguish the two sides, since $N$ moves away from one side and towards the other. Thus, one could attempt to formulate a weaker conjecture that applies only to one side. However a sensible conjecture should be time-reversal invariant~\cite{CEB1}. Under time-reversal, the putative weaker conjecture would require $N$ to move towards the opposite spatial side. But the left hand side of Eq.~(\ref{eq-sqfc}) is the same as if we had chosen that spatial side as the exterior with the original time direction, since $\Theta$ involves an even number of derivatives. This suggests that if there are any counterexamples to the QFC, then there will be counterexamples to a weaker conjecture that restricts attention to one side of $N$.

In the next two sections, we will provide evidence for the validity of the QFC. We will show that the QFC implies a {\em Quantum Bousso Bound}, for which there is already considerable evidence~\cite{CEB1,FMW,RMP,BouFla03,StrTho03,BCFM1,BCFM2}. We will also show that the QFC implies a previously unknown property of nongravitational theories, the {\em Quantum Null Energy Condition}, and we sketch a proof of this property.


\section{Quantum Bousso Bound}
\label{sec-qceb}

In this section, we will show that the QFC implies a quantum Bousso bound. For this purpose we will consider finite variations away from an initial surface $\sigma$ along the null hypersurface $N$ to some final surface $\sigma'$. We will take the surface $\sigma$ to correspond to $V(y)\equiv 0$; $\sigma'$ is defined by a choice of $V(y)\geq 0$ described below.

\subsection{The QFC Implies a Quantum Bousso Bound}
\label{sec-qfcceb}

Suppose that the quantum expansion at the generator $y_1$ is nonpositive (negative) on $\sigma$.  Then by integrating the QFC, we find that $\Theta$ will be nonpositive (negative) at $y_1$ at all later times: for any slice defined by a function $V(y)$, we have
\begin{equation}
\Theta[0, y_1] \le 0~,~ V(y)\geq 0 \implies \Theta[V(y),y_1] \le 0~,
\label{eq-wqfc}
\end{equation}
where if the first inequality is strict, so is the second.

Let us further specialize to the case where a later slice $\sigma'$ defined by $V(y)$ differs from $\sigma$ {\em only} on generators along which the generalized entropy is initially decreasing:
\begin{eqnarray} 
V(y) & \geq 0 & \mathrm{if}~\Theta[0,y]\leq 0\nonumber \\
V(y) & =0 & \mathrm{if}~\Theta[0,y]> 0~.
\label{eq-nonexp}
\end{eqnarray}
Eq.~(\ref{eq-wqfc}) implies that the generalized entropy decreases on these same generators on every intermediate slice $\alpha V(y)$, $0\leq \alpha\leq 1$.\footnote{The argument does not depend on how we interpolate between the initial and final slice, as long as the sequence of deformations is monotonic in the affine parameter. For example, we could begin by deforming $\sigma$ along one generator all the way, then along some other generator, etc., until the surface has been moved a distance $V(y)$ along each generator $y$.} Since the slice is deformed only along these generators, it follows that the generalized entropy must be less on $\sigma'$ than on $\sigma$:
\begin{equation}
S_\text{gen}[V(y)] \leq S_\text{gen}[0].
\label{eq-sls}
\end{equation}

To see that this implication is related to the Bousso bound, let us write out the result using Eq.~(\ref{eq-sgendef}):
\begin{equation}
S_\text{out}[\sigma']+\frac{A[\sigma']}{4G\hbar} \leq S_\text{out}[\sigma]+\frac{A[\sigma]}{4G\hbar}~,
\label{eq-sls2}
\end{equation} 
where we have left other counterterms implicit. Rearranging terms, we find 
\begin{equation}
S_\text{out}[\sigma^\prime] - S_\text{out}[\sigma]\leq \frac{A[\sigma]-A[\sigma^\prime]}{4G\hbar}~.
\label{eq-aass}
\end{equation}
Thus we recover the Bousso bound, Eq.~(\ref{eq-gceb}), if we identify
\begin{eqnarray} 
A[\sigma] & \equiv & A~,\\ 
A[\sigma^\prime] &\equiv & A'~, \\
S_\text{out}[\sigma^\prime] - S_\text{out}[\sigma] &\equiv & S~.\label{eq-sdifference}
\end{eqnarray}

However, it is important to note that the terms on the left and right hand side of Eq.~(\ref{eq-aass}) are separately cut-off dependent. Thus it is significant that the QFC yields the Bousso bound in the form of Eqs.~(\ref{eq-sls}) and (\ref{eq-sls2}), which are well-defined independently of a cutoff. 

The result is goes beyond the original Bousso bound, Eq.~(\ref{eq-gceb}), not only in that it clarifies how the matter entropy should be regulated for systems that are not well isolated, but more broadly in how $\Delta S$ should be defined.\footnote{However, in the weak gravity limit there exists an alternative, inequivalent regulator and definition of $\Delta S$. The corresponding quantum version of the Bousso bound was formulated and proven in~\cite{BCFM1,BCFM2}; see Sec.~\ref{sec-BCFM}.} Eq.~(\ref{eq-sdifference}) implies that the entropy cannot be determined from data on the null surface $N$ between $\sigma$ and $\sigma'$ alone. Instead, Eq.~(\ref{eq-sdifference}) instructs us to consider the von Neumann entropy on half-Cauchy-surfaces bounded by $\sigma$ and $\sigma'$, and compute their difference. Thus, data far from $N$ can affect the entropy.

As first noted by Strominger and Thompson~\cite{StrTho03}, the contributions of distant entanglement entropy are helpful in extending the validity of the Bousso bound into a regime where quantum effects on the metric are important, such as the evaporation of a black hole. For example, the Bousso bound in its original form would be violated on the horizon of a quantum black hole whose evaporation is sufficiently nearly balanced by an influx of entropic radiation~\cite{Low99}. 

Our Eq.~(\ref{eq-sls}), like the Strominger-Thompson proposal, evades this violation. Note that the condition of initial quantum nonexpansion, Eq.~(\ref{eq-nonexp}), is satisfied on the event horizon only if we consider past-directed light-sheets. Then Eq.~(\ref{eq-sls2}) is satisfied because the GSL is valid: the Hawking radiation increases the exterior von Neumann entropy by more than it decreases the Bekenstein-Hawking entropy of the black hole~\cite{Pag76}. The close relation of our result to the Strominger-Thompson proposal is discussed further in Sec.~\ref{sec-ST}.





\subsection{Recovering the Bousso Bound on Isolated Systems} 

The original Bousso bound is well-defined in the hydrodynamic regime, where entropy can be approximated as the integral of an entropy density. More generally, it is well-defined in the broad arena where an isolated matter system (e.g., a box of radiation) crosses the light-sheet $N$ between $\sigma$ and $\sigma'$. We assume that the matter system is well-separated from and unentangled with other matter systems, and also that the system is well-localized away from $\sigma$ and $\sigma^\prime$.  

\begin{figure}
	\centering
        
        \begin{subfigure}[b]{2.5in}
                \includegraphics[width=2.5in]{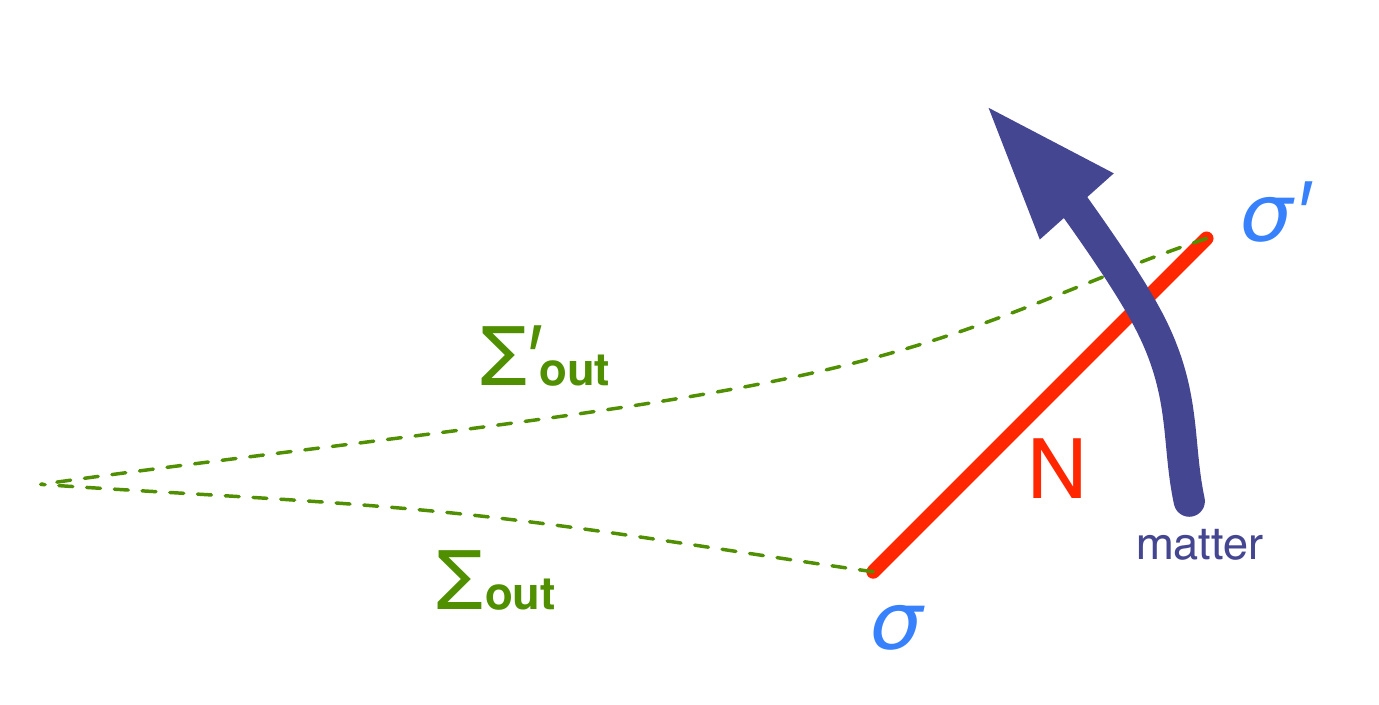}
                \caption{}
                \label{fig-nowife}
        \end{subfigure}
		\hspace{.4in}
        \begin{subfigure}[b]{2.5in}
                \includegraphics[width=2.5in]{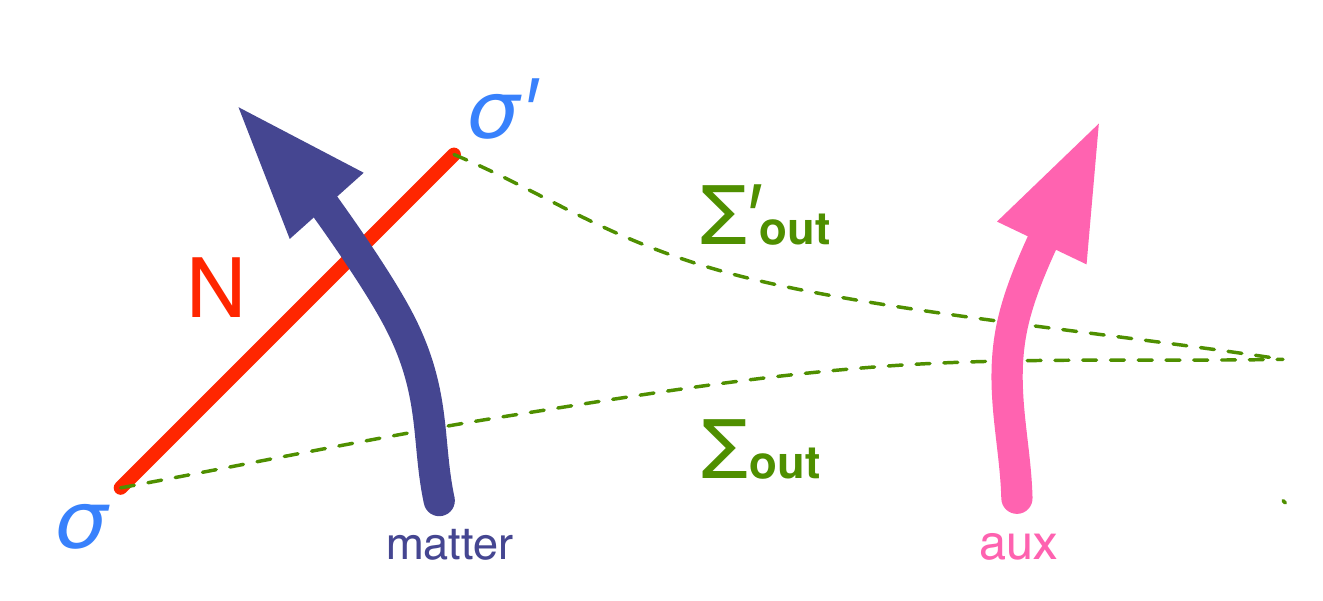}
                \caption{}
                \label{fig-nowife2}
        \end{subfigure}
        
        \caption{(a) For an unentangled isolated matter system localized to $N$, the quantum Bousso bound reduces to the original bound. (b) With the opposite choice of ``exterior,'' one can also recover the original entropy bound, by adding a distant auxiliary system that purifies the state.}
\end{figure}
Recall that we must choose one side of $N$ to define the exterior of the Cauchy surface ($\Sigma_\text{out}$), whose renormalized von Neumann entropy is used for computing $S_\text{gen}$. The Quantum Focussing Conjecture is valid regardless of which choice is made. For the purpose of recovering the Bousso bound, we choose, at $\sigma$, the spatial side opposite to $N$ (or equivalently, at $\sigma^\prime$, the same side as $N$). This is illustrated in Fig.~\ref{fig-nowife}. We make this choice independently of whether $\sigma$ lies in the future of $\sigma^\prime$ or vice-versa. 

With this choice, the exterior of $\sigma^\prime$ contains the same degrees of freedom as the union of $N$ with the exterior of $\sigma$. Moreover, with the above assumptions on the matter system(s) crossing $N$, the density operators factorize:
\begin{equation}
\rho_{\rm out}(\sigma') = \rho_{\rm out}(\sigma)\otimes \rho_N~,
\label{eq-rhofactor}
\end{equation}
where $\rho_N$ is the state of the isolated matter system. Hence the von Neumann entropies are additive:
\begin{equation} 
S_\text{out}[\sigma^\prime] =  S_\text{out}[\sigma] - \mathrm{tr}\, \rho_N \log \rho_N~.
\label{eq-sadd}
\end{equation}
Thus the vacuum entanglement entropy can be separated from the ``active'' matter entropy; the former can be regarded as already included in the geometric counterterms. Then Eq.~(\ref{eq-aass}) becomes the original Bousso bound
\begin{equation}
S \equiv - \mathrm{tr}\, \rho_N \log \rho_N \leq \Delta A/4G\hbar~, 
\label{eq-cebrec}
\end{equation}
with the entropy defined intrinsically as that of the isolated matter system(s) crossing the light-sheet $N$. 

There is more than one way to recover the Bousso bound for isolated systems. Suppose that in the above setting of an isolated system on the light-sheet $N$, we make the opposite choice of exterior; see Fig.~\ref{fig-nowife2}. This has the effect of exchanging $\sigma$ and $\sigma'$ in Eqs.~(\ref{eq-rhofactor}) and (\ref{eq-sadd}). Instead of Eq.~(\ref{eq-cebrec}) we obtain the bound $-S<\Delta A/(4G\hbar)$, where $S=-\mathrm{tr}\, \rho_N\log \rho_N$ is the entropy of the isolated system. This bound is valid but not very interesting: it is trivially satisfied for ordinary matter systems, since $S>0$ and $\Delta A>0$ in the classical limit. 

However, let us now add an auxiliary system to the exterior (far from $N$) that purifies the mixed state $\rho_N$ of the matter crossing $N$. The initial entropy $S[\sigma]$ receives no contribution from the matter and auxiliary systems because they are both present and in a pure state. But the final entropy $S[\sigma']$ is just that of the purification, and hence equal to that of the matter system, $S=-\mathrm{tr}\, \rho_N\log \rho_N$. Thus, we recover Eq.~(\ref{eq-cebrec}): with the inclusion of a distant purification, the bound on the isolated matter system is again nontrivial and equivalent to the previous example.

\subsection{The Role of Quantum Nonexpansion} 
\label{sec-nonexp}

As originally formulated, the Bousso bound applies only to light-sheets, i.e., to null hypersurfaces whose classical expansion $\theta$ is nonpositive everywhere in the direction from $\sigma$ to $\sigma'$, and which contain no caustics. (It follows that $A-A'\geq 0$.) 
However, the classical nonexpansion assumption played no role in our formulation of the quantum Bousso bound.

In fact, classical nonexpansion can be violated in settings where our bound applies. Consider a slice $\sigma$ of the event horizon of an evaporating black hole. The generalized entropy is decreasing towards the past, so the conjecture applies if $\sigma^\prime$ is chosen as an {\em earlier} horizon slice. But then $A[\sigma^\prime]>A[\sigma]$, so the classical expansion is positive.

Instead, our quantum Bousso bound substitutes quantum expansion for classical expansion: we restricted to deformations along the generators for which the initial {\em quantum expansion} at $\sigma$ is nonpositive. In this sense, Eq.~(\ref{eq-nonexp}) can be taken as the definition of a {\em quantum light-sheet}. 

The QFC then plays an interesting dual role. First, it guarantees that the quantum expansion is nonpositive not only initially, but everywhere on $N$ between the two slices $\sigma$ and $\sigma^\prime$. And second, the resulting non-positivity of its integral (the difference between the final and initial generalized entropies) becomes the statement of the entropy bound.

The classical Bousso bound can formulated in an alternate way~\cite{CEB1}, more closely analogous to the quantum version we have constructed. Instead of demanding that $\theta\leq 0$ everywhere on a light-sheet, one could have demanded classical nonexpansion only initially, but assumed the null energy condition, and added a requirement that light-sheets must be terminated at caustics.  The role of the null energy condition would then be to ensure that $\theta\leq 0$ everywhere on the null surface, by the classical focussing theorem. However, the bound itself would still be a separate statement; it does not also follow from the null energy condition.

\section{Quantum Null Energy Condition}
\label{sec-qnec}

In the previous section, we considered the integrated QFC and showed that it implies a quantum Bousso bound. The considerable evidence for the Bousso bound thus supports the QFC. 

We now return to the QFC as a constraint on a second functional derivative,
\begin{equation}
\frac{\delta}{\delta V(y_2)}  \frac{4G\hbar}{\sqrt{^V\!g(y_1)}} \frac{\delta S_\text{gen}}{\delta V(y_1)}~ \le 0~,
\end{equation}
and we examine whether there exist other limits in which one can find explicit evidence or formulate a proof of the conjecture. We begin in Sec.~\ref{sec-offdiagonal} with a proof of the QFC in the off-diagonal case, $y_1\neq y_2$. In Sec.~\ref{sec-diagonal} we consider the diagonal part, $y_1=y_2$. We show that it gives rise to an interesting limit when the classical null extrinsic curvature vanishes: the Quantum Null Energy Condition, a nongravitational implication of the QFC.
In Sec.~\ref{sec-proof} we outline a proof of this field theoretic statement.

\subsection{General Proof of the Off-diagonal QFC}
\label{sec-offdiagonal}

\begin{figure}

	\centering

        \begin{subfigure}[b]{2.5in}
                \includegraphics[width=2.5in]{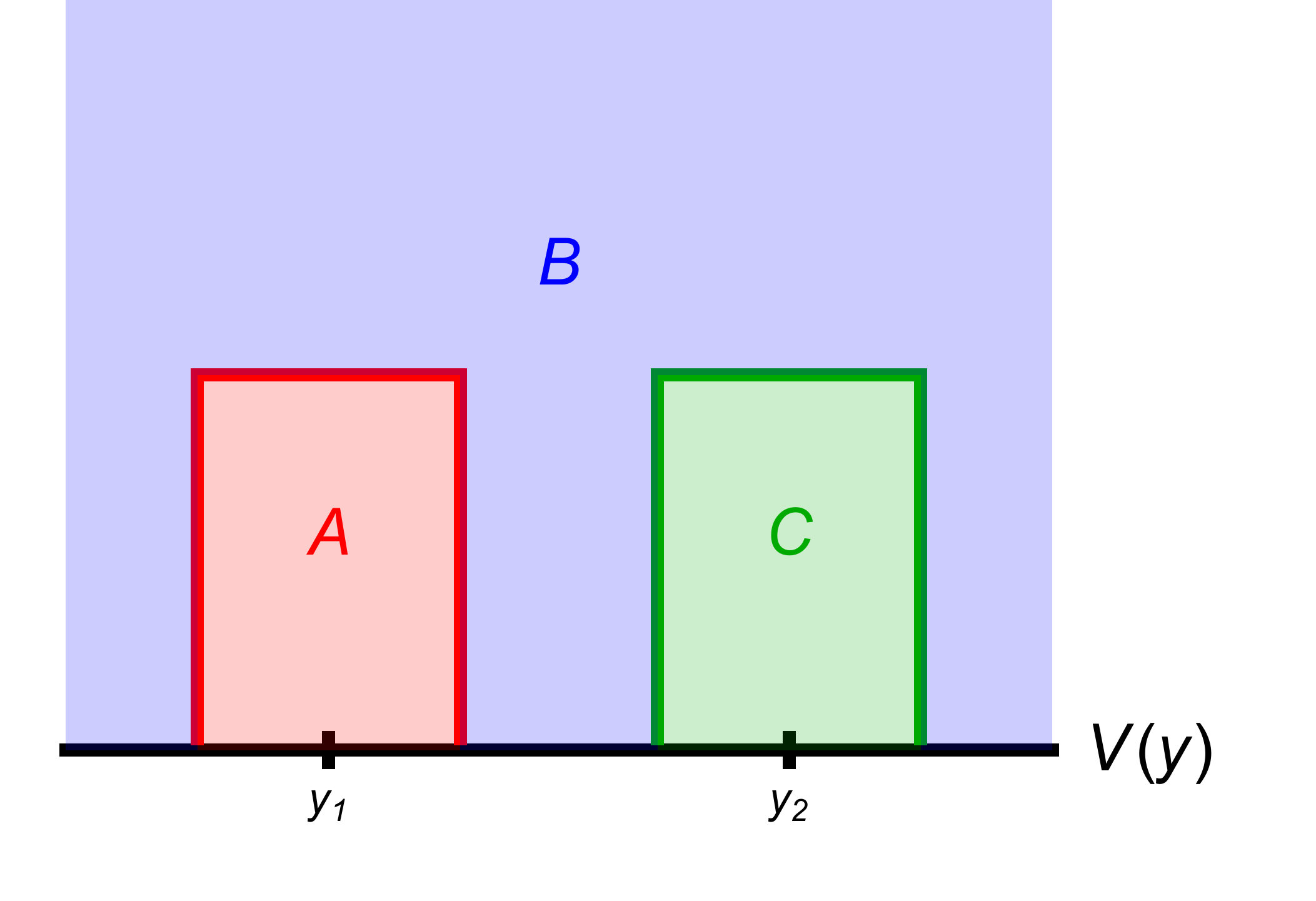}
                \caption{}
                \label{fig-offdiagonal}
        \end{subfigure}
		\hspace{.4in}
        \begin{subfigure}[b]{2.5in}
                \includegraphics[width=2.5in]{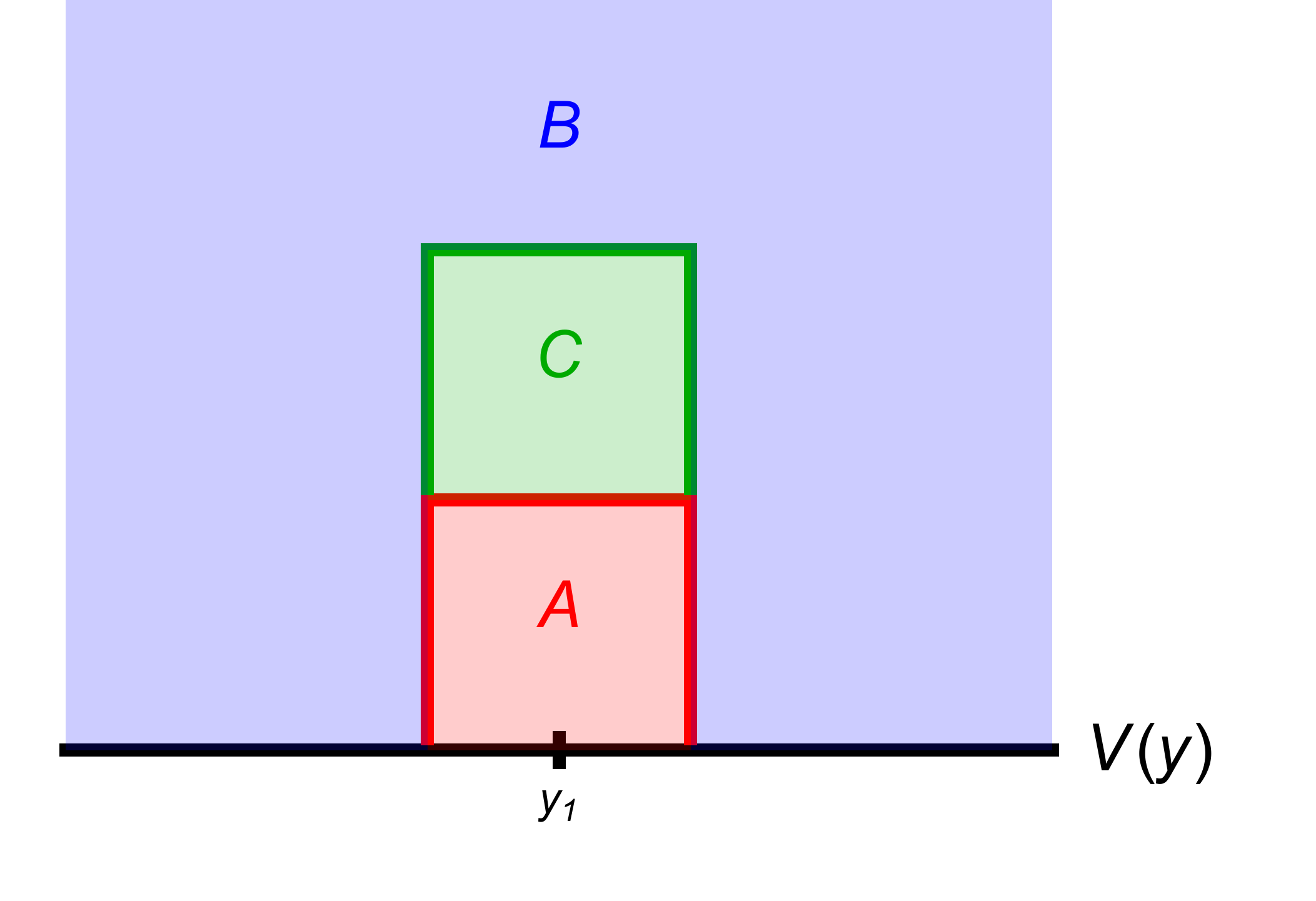}
                \caption{}
                \label{fig-diagonal}
        \end{subfigure}

	\caption{(a) A portion of the null surface $N$, which we have chosen to coincide with $\Sigma_{\rm out}$ in the vicinity of the diagram. The horizontal line at the bottom is the surface $V(y)$, and the red and green lines represent deformations at the transverse locations $y_1$ and $y_2$. The region above both deformations is the region outside of $V_{\epsilon_1,\epsilon_2}(y)$ and is shaded blue and labeled $B$. The region between $V(y)$ and $V_{\epsilon_1}(y)$ is labeled $A$ and shaded red. The region between $V(y)$ and $V_{\epsilon_2}(y)$is labeled $C$ and shaded green. Strong subadditivity applied to these three regions proves the off-diagonal QFC.
	(b) A similar construction for the diagonal part of the QFC. In this case, the sign of the second derivative with respect to the affine parameter is not related to strong subadditivity.
	}

\end{figure}

For $y_1 \ne y_2$, the QFC follows from strong subadditivity. Since $A[V(y)]$ is the integral of a local functional of $V(y)$, and the factor $\sqrt{^V\!g}$ appearing the QFC is evaluated at $y_1$, it follows that the off-diagonal second derivative only receives a contribution from $S_\text{out}$:
\begin{equation}
\frac{\delta}{\delta V(y_2)}  \frac{4G\hbar}{\sqrt{^V\!g(y_1)}} \frac{\delta S_\text{gen}}{\delta V(y_1)} = \frac{4G\hbar}{\sqrt{^V\!g(y_1)}} \frac{\delta^2 S_\text{out}}{\delta V(y_1)\delta V(y_2)} ~~~\text{for } y_1 \ne y_2.
\end{equation}
The functional derivative can be realized as the following limit:
\begin{equation}\label{eq-discreteoffdiag}
\begin{split}
&\frac{\delta^2 S_\text{out}}{\delta V(y_1)\delta V(y_2)} = \\
&\lim_{\mathcal{A}_1,\mathcal{A}_2\to 0}  \frac{\sqrt{^V\!g(y_1)^V\!g(y_2)}}{\mathcal{A}_1\mathcal{A}_2} \lim_{\epsilon_1,\epsilon_2\to 0}  \frac{S_\text{out}[V_{\epsilon_1,\epsilon_2}(y)]-S_\text{out}[V_{\epsilon_1}(y)]-S_\text{out}[V_{\epsilon_2}(y)]+S_\text{out}[V(y)]}{\epsilon_1\epsilon_2}.
\end{split}
\end{equation}
Here $\mathcal{A}_i$ is an area element located at $y_i$, and $\epsilon_i$ is a deformation parameter for the surface along the generator at $y_i$. To be precise, we define
\begin{equation}
V_{\epsilon_1,\epsilon_2}(y) \equiv V(y) + \epsilon_1 \vartheta_{y_1}(y) + \epsilon_2 \vartheta_{y_2}(y),
\end{equation}
where $\vartheta_{y_i} =1$ in a neighborhood of area $\mathcal{A}_i$ around $y_i$. For brevity of notation, when $\epsilon_i =0$ we omit it from the subscript of $V_{\epsilon_1,\epsilon_2}(y)$. The relevant surfaces are depicted in Fig.~\ref{fig-offdiagonal}. It is clear from the figure that the numerator in Eq.~\ref{eq-discreteoffdiag} is negative by strong subadditivity. Specifically, we split the region outside of $V(y)$ into three subregions.\footnote{The entropies $S_{\rm out}[V(y)]$ are defined in terms of half-Cauchy surfaces $\Sigma_{\rm out}$, and for the purposes of this discussion we are using the freedom to unitarily deform $\Sigma_{\rm out}$ so that it lies along the null surface $N$. Thus the subregions $A$, $B$, and $C$ which we define are subregions of $N$.} The subregion outside of $V_{\epsilon_1,\epsilon_2}(y)$ will be called region $B$, so that $S_\text{out}[V_{\epsilon_1,\epsilon_2}(y)] = S(B)$. The subregion outside of $V_{\epsilon_1}(y)$ but inside of $V_{\epsilon_1,\epsilon_2}(y)$ will be called $C$, so that $S_\text{out}[V_{\epsilon_1}(y)] = S(BC)$. Subregion $A$ is defined in a similar way so that $S_\text{out}[V_{\epsilon_2}(y)] = S(AB)$. Finally, we have $S_\text{out}[V(y)] = S(ABC)$. In this notation, the numerator of Eq.~\ref{eq-discreteoffdiag} is the standard combination of entropies appearing in the strong subadditivity inequality:
\begin{equation}
S(B) - S(BC) - S(AB) + S(ABC) \le 0~.
\end{equation}
This is enough to prove the off-diagonal QFC in general.

We also would like to emphasize that the combination of entropies appearing here is finite and cutoff-independent. This has to be the case because the generalized entropy we started with was cutoff-independent, but it is instructive to see this directly at the level of the matter entropy. Sometimes in quantum field theory, entropy inequalities are true because of cutoff-dependent terms, and in the continuum limit reduce to the trivial statement $-\infty < 0$. Here that is not the case. The cutoff-dependent terms in the entanglement entropy for a given region are proportional to integrals of geometric quantities along the boundaries of that region. One can check that, for the regions $A$, $B$, and $C$ that we have defined, such terms cancel in the combination $S(B) - S(BC) - S(AB) + S(ABC)$.

Our construction here is similar to the ``entanglement density'' of \cite{Bhattacharya:2014vja, Lashkari:2014kda}, although the QFC is stronger in that it also places a constraint on the diagonal terms with $y_1 = y_2$.  Strong subadditivity is not helpful in this case: if we attempted to use the same strategy, then (with appropriately modified definitions of the subregions, see Fig.~\ref{fig-diagonal}), we would find a combination of entropies $S(B) - 2S(BC)+ S(ABC)$, which does not have any direct relation with strong subadditivity.  Furthermore, for this combination of entropies the cutoff-dependent terms do not cancel.  This is no surprise since the diagonal part of the QFC receives contributions from the area term, and the area term is essential for making $S_{\rm gen}$ cutoff-independent in general.

\subsection{Diagonal Part of the QFC}
\label{sec-diagonal}

The case $y_1=y_2$ corresponds to a deformation of the surface $\sigma$ along a single null generator orthogonal to it. Specializing to this case, it is convenient to work with ordinary derivatives with respect to the affine parameter along generator $y_1$, denoted by primes. By Eq.~(\ref{eq-sgendef})  we have
\begin{equation}\label{qtheta}
\Theta = \theta + \frac{4G\hbar}{\mathcal A} S'_\text{out}~,
\end{equation}
where $\theta$ is the classical expansion.\footnote{Here and in the remainder of this section we will not explicitly write ${\rm lim}_{{\cal A}\to 0}$ in our expressions, but it should always be understood.} 
 The QFC becomes
\begin{eqnarray} 
0\geq \Theta' & = & \theta' + \frac{4G\hbar}{\mathcal A} \left(S''_\text{out}-S_\text{out}'\theta\right)\\
& = & -\frac{1}{2}\theta^2 -\varsigma^2 -8\pi G \langle T_{kk}\rangle + \frac{4G\hbar}{\mathcal A} \left(S''_\text{out}-S_\text{out}'\theta\right)
\end{eqnarray} 
The derivatives of $S_{\rm out}$ scale linearly with $\cal{A}$, matching the scaling of the other terms. Any terms that go like higher powers of ${\cal A}$ will drop out as ${\cal A}\to 0$. 

The above form shows that the QFC has several interesting limits. The most obvious is the classical limit, $\hbar\to 0$. In this case, one recovers the null energy condition, $\langle T_{kk}\rangle \geq 0$, which must hold since at any point $p$ one can consider a congruence with tangent vector $k^a$, such that the shear $\varsigma$ and the expansion $\theta$ both vanish at $p$.
\footnote{When $\hbar\to 0$, quantum corrections to $\langle T_{kk}\rangle$ proportional to $\hbar$ vanish. The null energy condition is recovered for the $\hbar^0$ term in the semiclassical expansion of $\langle T_{kk}\rangle$.} 
For arbitrary congruences it follows that $\theta'\leq 0$.

The same special choice of congruence, with $\theta=\varsigma=0$, becomes more interesting if we do not set $\hbar\to 0$. In this case the QFC implies the relation
\begin{equation}
\langle T_{kk} \rangle \geq \frac{\hbar}{2\pi\mathcal A} S''_\text{out}~,
\label{eq-qnec}
\end{equation}
which we shall call the {\em Quantum Null Energy Condition} (QNEC).  

It is extremely intriguing that \eqref{eq-qnec} does not depend on $G$.  Nor is it affected by higher curvature terms in the gravitational action, at least when the stationary null congruence is also a Killing horizon.  (As discussed in Appendix~\ref{ren}, these higher order terms arise due to quantum loop corrections, and add additional terms to the gravitational entropy $S_\text{grav}$, besides the area.  This results in modifications to \ref{qtheta}.  However, the gravitational equations of motion also change in exactly the right way \cite{Wall:2015raa, Bhattacharjee:2015yaa} so that, for linearized metric perturbations to the Killing horizon, $S''_\text{grav} = 2\pi \langle T_{kk} \rangle$.  Hence the form of \ref{eq-qnec} remains the same.)

Thus the QNEC is entirely a statement about quantum field theory.  It is the effective quantum replacement for the null energy condition, and unlike the null energy condition it is something that might follow from first principles in quantum field theory.

As a first nontrivial check, note that the QNEC is satisfied by an infinite class of states in any 1+1 CFT, namely those which are conformally related to the vacuum state (or to coherent states in a Gaussian theory). This follows from the anomalous transformation properties of $T_{kk}$ and $S_\text{out}$ under a general conformal transformation, if we note that $S''_\text{out}=0$ and $T_{kk}\geq 0$ for vacuum/coherent states on a causal horizon~\cite{Wall:2011kb}.\footnote{In a theory with 1+1 conformal symmetry the QNEC in fact implies a slightly stronger statement, namely $\frac{2\pi}{\hbar} T_{kk} - S''_\text{out} - \frac{6}{c} (S'_\text{out})^2 \ge 0$, where $c$ is the central charge (and ${\mathcal A} = 1$ since in two dimensions $\sigma$ is a point).  That is because this quantity transforms as a primary under conformal transformations, and one can always find a conformal frame where $S'_\text{out} = 0$ \cite{Wall:2011kb}.}  Indeed, it was suggested in~\cite{Wall:2011kb} that what we here call the QNEC might hold for more general states and in higher dimensions.

A proof of the QNEC in the case of free field theory and causal horizons will be presented in forthcoming work.  We will now outline the strategy of this proof. 

\subsection{Sketch of Proof in the Weak Gravity Limit}
\label{sec-proof}

The essential features of the Quantum Null Energy Condition are illuminated if we consider perturbations around flat Minkowski space, where $\sigma$ is a codimension-2 plane and the null surface $N$ is a Rindler horizon (a codimension-1 null hyperplane).

A helpful insight comes from null quantization, which is valid for free theories,\footnote{or superrenormalizable theories} and so we restrict ourselves to free theories in the following. The entropy $S_\text{out}$ refers to the entropy on a spacelike Cauchy surface, but by unitary time evolution we can alternatively think of it as the entropy of the state restricted to the part of $N$ in the future of $\sigma$, together with a portion of null infinity. We are considering deformations of $\sigma$ along a single generator at location $y$, and the amount of the deformation is determined by an affine parameter $\lambda >0$. We will ``thicken'' the generator by deforming the surface in a transverse area ${\cal A}$ around the generator, and the thickened generator is called the pencil. At the end we take ${\cal A}\to 0$, and in that limit we should recover \eqref{eq-qnec}.

We will use the remarkable fact that the vacuum state restricted to a causal horizon is actually a product state over the generators of the horizon, and each factor looks like the vacuum state of a 1+1-dimensional 
chiral CFT \cite{Wall11}. 
Our global state is not the vacuum, of course, but in the limit ${\cal A}\to 0$ the density matrix on the pencil should 
approach that of the vacuum, for the usual reason that all finite energy states look like the vacuum in the limit of a very small region. Restricting the global state to the region in the future of $\sigma$ means taking a partial trace over the region to the past of $\sigma$. Increasing the affine parameter $\lambda$ along the pencil corresponds to taking additional partial traces along the pencil subsystem, while leaving the other generators alone. We will lump all of the other generators (together with a portion of null infinity) into a single unit called the auxiliary system. All of these facts can be summarized in the equation the for density matrix of the system,
\begin{equation}
\rho(\lambda) = \rho^{(0)}_{\rm pen}(\lambda) \otimes \rho^{(0)}_{\rm aux} + \delta\rho(\lambda).
\end{equation}
Here $\rho^{(0)}_{\rm pen}(\lambda)$ is the vacuum state reduced density matrix on the part of the pencil with affine parameter greater than $\lambda$, and $\rho^{(0)}_{\rm aux}$ is an arbitrary state in the auxiliary system. The perturbation $\delta\rho(\lambda)$ contains entanglement between the auxiliary system and the pencil, and is small: we will argue below that it is proportional to ${\cal A}^{1/2}$.  Therefore we will treat $\delta\rho(\lambda)$ as an expansion parameter in our computation. We will assume that $\rho^{(0)}(\lambda) \equiv \rho^{(0)}_{\rm pen}(\lambda) \otimes \rho^{(0)}_{\rm aux}$ is a properly normalized density matrix, so ${\rm tr(\delta\rho) = 0}$.

$S_\text{out}(\lambda)$ is the von Neumann entropy of $\rho(\lambda)$, and we will now compute it as a perturbation series in $\delta\rho$:
\begin{equation}\label{eq-Slambda}
S_\text{out}(\lambda) = - {\rm tr}\left[\rho^{(0)}(\lambda) \log \rho^{(0)}(\lambda)\right]  - {\rm tr}\left[\delta\rho(\lambda) \log \rho^{(0)}(\lambda)\right] +O(\delta\rho^2).
\end{equation}
The $\delta\rho$-independent term appearing here, which  we will call $S_\text{out}^{(0)}$, actually does not depend on $\lambda$. To see this, we use the fact that $\rho^{(0)}$ is a product state, so its logarithm is a sum of two terms. Then we have
\begin{equation}
S_\text{out}^{(0)} \equiv - {\rm tr}\left[\rho^{(0)}(\lambda) \log \rho^{(0)}(\lambda)\right] =  - {\rm tr}\left[\rho_{\rm pen}^{(0)}(\lambda) \log \rho_{\rm pen}^{(0)}(\lambda)\right]- {\rm tr}\left[\rho_{\rm aux}^{(0)} \log \rho_{\rm aux}^{(0)}\right].
\end{equation}
The second term on the right-hand side is manifestly independent of $\lambda$, and the first is independent of $\lambda$ because the vacuum state on the pencil is invariant under translations in the affine parameter.\footnote{More precisely, the reduced density matrices $\rho^{(0)}_{\rm pen}(\lambda)$ for different values of $\lambda$ are related to each other by unitary transformations, namely translations in $\lambda$.}

We may perform a similar decomposition on the term in \eqref{eq-Slambda} linear in $\delta\rho$, which we label $\Delta K$:
\begin{equation}
\Delta K(\lambda) \equiv -{\rm tr}\left[\delta\rho(\lambda) \log \rho^{(0)}(\lambda)\right] = -{\rm tr}\left[\delta\rho(\lambda) \log \rho_{\rm pen}^{(0)}(\lambda)\right] -{\rm tr}\left[\delta\rho(\lambda) \log \rho_{\rm aux}^{(0)}\right].
\end{equation}
By evaluating the trace over the pencil subsystem in the second term, we can see that it is actually $\lambda$-independent. To identify the first term, we use the fact that $\rho^{(0)}_{\rm pen}(\lambda)$ is just the Rindler density operator for the 1+1-dimensional CFT on the pencil, so in particular it is thermal with respect to the Rindler boost generator. Then we have the identity
\begin{equation}
-{\rm tr}\left[\delta\rho(\lambda) \log \rho^{(0)}_{\rm pen}(\lambda)\right] = \frac{2\pi {\cal A}}{\hbar} \int_\lambda^\infty dx  \,(x-\lambda)\langle T_{kk}(x)\rangle,
\end{equation}
where the integral is along the generator at $y$. In particular, taking two derivatives of this expression with respect to $\lambda$ exactly produces the energy-momentum term appearing in \eqref{eq-qnec}.

To summarize, we have shown the $\lambda$-dependence of the terms in \eqref{eq-Slambda} implies the equation 
\begin{equation}
(\Delta K - S_\text{out} + S_\text{out}^{(0)})'' = \frac{2\pi {\cal A}}{\hbar}\langle T_{kk}\rangle - S_\text{out}''.
\end{equation}
Therefore, the QNEC \eqref{eq-qnec} reduces to the statement that the $O(\delta \rho^2)$ terms appearing in the expansion of $S_\text{out}$ have a negative second derivative in the limit ${\cal A}\to 0$. Notice that we only need to worry about terms in the expansion of $S_\text{out}$ which scale linearly with ${\cal A}$; terms which vanish more quickly will drop out in the limit. Earlier we claimed that $\delta \rho$ was proportional to ${\cal A}^{1/2}$, which means that it is only the $\delta \rho^2$ term in the expansion of $S_\text{out}$ which contributes. We will now argue for this scaling.

Consider the state on the full pencil, $\lambda \to -\infty$, so that nothing on the pencil has been traced out. Then the vacuum state is the zero-particle Fock state, $|0\rangle\!\langle 0|$. Suppose we expand $\delta\rho$ on the full pencil in the Fock basis. If $|m\rangle$ is an $m$-particle state, then the coefficient of $|m\rangle\!\langle m|$ in the expansion of $\delta\rho$ is expected to scale like ${\cal A}^m$, simply because that is the scaling of the probability to measure $m$ particles in a small volume (remember that we first choose the state and then take $\cal{A}$ to be small). Then positive definiteness of the total density matrix implies that the largest terms in $\delta\rho$ are of the form $|0\rangle\!\langle 1|$ and $|1\rangle\!\langle 0|$, and have coefficients which scale like ${\cal A}^{1/2}$. This means that $\delta\rho^2$ has terms of order $\cal{A}$, but higher powers of $\delta\rho$ contain higher powers of ${\cal A}$. This structure is preserved as we take traces.

We have shown that the QNEC reduces to the statement that the term in the expansion of $S_\text{out}$ which is second order in $\delta \rho$ has a negative second derivative. We have effectively removed the geometry from the problem: it is enough to consider a 1+1-dimensional free chiral CFT entangled with an arbitrary auxiliary system. Using the replica trick, for instance, one may prove the statement by calculating the von Neumann entropy explicitly. A proof along these lines is the subject of a forthcoming paper.

It is also intriguing to consider a generalization of the QNEC beyond field theory. The quantity $\Delta K - S_\text{out} + S_\text{out}^{(0)}$ is usually called the relative entropy:
\begin{equation}
S(\rho || \rho^{(0)}) \equiv {\rm tr} \left[\rho \log \rho -\rho \log \rho^{(0)}\right].
\end{equation}
The relative entropy between any two states of an arbitrary quantum system is a measure of their distinguishability, and is a quantity of particular significance in quantum information theory (see \cite{Ved02} for a general review, and \cite{Cas08} for the use of relative entropy in the gravitational context). So we can attempt to generalize the QNEC to other quantum systems as the statement that $S(\rho||\rho^{(0)})$ has a non-negative second derivative with respect to $\lambda$, where $\lambda$ parametrizes a chosen coarse-graining operation on the pair of states (or that such a statement holds at least for the term quadratic in $\delta\rho$). The first derivative with respect to $\lambda$ is known to be non-positive for all pairs of states in any quantum system and for any choice coarse-graining operation: this is the famous monotonicity property of relative entropy. We do not know of a general result regarding the second derivative, and it would be interesting to identify the class of quantum systems, states, and coarse-graining operations for which it is non-negative.\footnote{In the context of a system governed by Boltzmann's equations, \cite{Garrett} showed that the first $n$ derivatives of the entropy all alternate in sign, for at least $n \le 6$ (but not for all $n$).  But in our case, it is clear that there is no constraint on the sign of the third or higher derivatives of $S_\text{gen}$, since in the classical regime the derivatives of $T_{kk}$ can take any sign.}

\section{Relationship to Other Conjectures and Results}
\label{sec-other}

In this section, we discuss how our conjecture and its implications are related to older conjectures and results.

\subsection{Generalized Second Law for Causal Horizons}
\label{sec-GSL}

The generalized second law (GSL) states that the generalized entropy of a causal horizon is non-decreasing. The QFC can be applied more broadly to any surface that splits a Cauchy surface. But in particular, the QFC can be applied to cross-sections of a causal horizon, and it is natural to ask how it relates to the GSL in this setting.

A key difference is that the GSL constrains the sign of the first derivative of the generalized entropy, while the QFC constrains the sign of the second derivative. Thus it is clear that the conjectures are not equivalent on causal horizons.

However, the conjectures are related. Assuming that the GSL holds at one time, integrating the QFC implies that the GSL holds at all earlier times on the same causal horizon. Moreover, for causal horizons at late times, the classical expansion $\theta$ vanishes, and one expects the matter entropy $S_\text{out}$ to stop evolving. Thus, one expects that $\Theta\to 0$ in the asymptotic future for a (future) causal horizon. With this assumption, the QFC implies the GSL on the entire causal horizon.

\subsection{Strominger-Thompson Quantum Bousso Bound}
\label{sec-ST}

Strominger and Thompson \cite{StrTho03} proposed adding the ``entanglement entropy across the surface'' to the area of any cross-section of a light-sheet, to obtain a quantum Bousso bound. In particular, it was noted that the leading divergences in the entanglement entropy are cancelled by a renormalization of Newton's constant. The quantum bound we derive in Sec.~\ref{sec-qceb} automatically inherits these important features from the QFC. Thus we largely reproduce the Strominger-Thompson proposal as a special case of the QFC.

Our formulation of the quantum entropy bound differs in that we consider the generalized entropy as a fundamental object. Hence we do not distinguish between ``entanglement entropy'' across $\sigma$, and the entropy of other matter outside the surface $\sigma$. The latter is treated as a separate contribution in~\cite{StrTho03}, in a hydrodynamic approximation~\cite{FMW,BouFla03}. In general, the distinction between gravitational entropy, entanglement entropy, and ``matter entropy'' is ambiguous. 

By referring only to the generalized entropy, we were able to sidestep this ambiguity. Moreover, the generalized entropy regulates not only the leading (area) divergence of the entanglement entropy, but also subleading divergences such as the logarithmic divergence in $3+1$ spacetime dimensions.

The use of a hydrodynamic approximation also entered into the nonexpansion condition that defines valid light-sheets in the presence of matter~\cite{StrTho03}. Here this condition is universally given by Eq.~(\ref{eq-nonexp}): a light-sheet is generated by orthogonal light-rays with initially nonpositive {\em quantum expansion} $\Theta$.

It is interesting that in the hydrodynamic limit, the Quantum Null Energy Condition, Eq.~(\ref{eq-qnec}), reduces to one of the assumptions that underly the proof of the quantum Bousso bound, ($\partial_+s_+\leq 2T_{++}$ in the notation of~\cite{StrTho03}, a weakened version of an assumption introduced in~\cite{BouFla03}).

\subsection{BCFM Quantum Bousso Bound}
\label{sec-BCFM}

In the weak gravity limit, $G\hbar\to 0$
and $G(\langle T_{\mu\nu}\rangle - T_{\mu\nu}^\text{vac}) \to 0$,
one can restrict both the vacuum and the state of interest to the same region or light-sheet. Then a {\em vacuum-subtracted entropy} $\Delta S$ can be defined as the difference between the von Neumann entropies of the state and the vacuum~\cite{HolLar94,MarMin04,Cas08}. Because the divergences of the entanglement entropy are associated with its boundary, this quantity is finite and reduces to the expected entropy for isolated systems and fluids.\footnote{In the interacting case and on a light-sheet, it reduces to an upper bound on the na\"ive entropy, which suffices.} With this definition, a quantum Bousso bound can be proven to hold on any portion of a light-sheet~\cite{BCFM1,BCFM2}.

When gravitational backreaction of the state is not negligible, the spacetime geometry is very different from that of a vacuum state.  Then it is unclear what one would mean by restricting both a general state and the vacuum to the ``same'' region or light-sheet. In this case, one cannot define a finite entropy by vacuum subtraction.  

Here, we use a different method to regulate the divergence of the von Neumann entropy of quantum fields in a bounded region: we combined the matter entropy with the gravitational entropy to obtain a cutoff-independent, generalized entropy. This definition requires a semi-classical regime, but not that the gravitational backreaction is small. Therefore, the QFC does not require gravity to be weak; and the associated quantum formulation of the Bousso bound, too, can be stated in settings where gravity is strong. 

If gravitational backreaction is small, both statements can be applied. This is interesting, because they appear to be inequivalent. In~\cite{BCFM1,BCFM2}, the entropy is defined intrinsically on the light-sheet portion of interest, with no reference to distant spatial regions. The generalized entropy, by contrast, generically depends on regions far from the light-sheet. 

In fact, the light-sheets themselves are defined differently. The BCFM bound requires that a light-sheet have nonpositive {\em classical} expansion {\em everywhere} on $L$. Our bound requires that the {\em quantum} expansion be nonpositive {\em initially} on $L$; the QFC then becomes the {\em statement} of the entropy bound.


\subsection{Quantum Singularity Theorem}
\label{sec-singularity}

Penrose's singularity theorem~\cite{Pen65} is a seminal result in general relativity. It states that, in a globally hyperbolic spacetime which satisfies the null energy condition, that the presence of certain compact surfaces $T$ on a connected, noncompact Cauchy slice $\Sigma$ indicates that the spacetime is null geodesically incomplete. The surfaces which signal the impending breakdown of the spacetime are \textit{trapped surfaces}: surfaces for which the congruence of outgoing null lightrays have everywhere negative expansion. The proof uses the Raychaudhuri equation to argue that these null geodesics must reach a caustic in finite affine parameter. If the spacetime were null geodesically complete, then each null generator includes its endpoints. 

Because of the assumption of the null energy condition, Penrose's theorem is not applicable to quantum matter. Interestingly, there exists a generalization of Penrose's theorem, where the role of the area is replaced with the generalized entropy \cite{Wall10}. A \textit{quantum trapped surface} is defined in a globally hyperbolic spacetime as follows. Suppose that on some Cauchy surface $\Sigma$, a compact codimension-2 surface $\mathcal T$ exists, and its exterior is non-compact. If $N$ is the null surface generated by outward future-directed light rays, and if the generalized entropy is decreasing with time with respect to future null deformations, then $\mathcal T$ is called a quantum trapped surface. In the classical limit, the generalized entropy is simply the area, so this criteria reduces to the classical notion of a trapped surface.

The quantum proof~\cite{Wall10} is similar to the classical one. One starts with the assumption of a non-compact Cauchy surface containing a quantum trapped surface. Unlike the classical case which required the null energy condition, one now assumes that the generalized second law holds (i.e., the generalized entropy cannot decrease on causal horizons). The GSL implies (by contradiction) that the null generators reach caustics in finite affine parameter time. From here, the proof is proceeds as in the classical case: the non-compact surface $\Sigma$ cannot evolve into a compact surface, which implies that the endpoints of the null geodesics do not belong to the spacetime.

The QFC was not necessary to complete the proof of the quantum singularity theorem, but it does have interesting consequences for quantum trapped surfaces which makes them more analogous to their classical counterparts. For example, the outgoing null rays from a classical trapped surface define in an obvious way a sequence of additional trapped surfaces on Cauchy slices to the future. This result only becomes valid in the quantum case under the assumption of the weak quantum focussing theorem. 

\subsection{Barriers to Quantum Extremal Surfaces}
\label{sec-barrier}

The prescription for holographic entanglement entropy is now fairly well-understood at leading order in $1/N$ in the Anti-de Sitter/Conformal Field Theory correspondence \cite{Mal97}. An ample body of evidence supports the proposal \cite{HubRan07} of Hubeny, Rangamani, and Takayanagi, which extends an earlier proposal \cite{RyuTak06} by Ryu and Takayanagi. The new proposal is that, to calculate the entanglement entropy of a region $R$ of a CFT, we need to find a codimension-2 surface $X$ such that $\partial X = \partial R$, and $X$ homologous to $R$, which extremizes the area functional. If there are many such surfaces, we are instructed to pick a surface with the minimum area. The entanglement entropy of $R$ is then the area of $X$, in (bulk) Planck units: $S_R = A_X/4 G \hbar$. Since $G\hbar \sim 1/N^2$, this conjecture gives the leading order entropy in a $1/N$ expansion, but as in the black hole case, there will generally be subleading corrections.  

Recently, the next-to-leading order corrections in $1/N$ to the entanglement entropy were calculated. The proposal \cite{FLM13} of Faulkner, Lewkowycz and Maldacena (FLM) is to take the leading order prescription, that is to calculate the area of an extremal surface in the bulk, and to add in the von Neumann entropy of the bulk state restricted to one side of the extremal surface. That is, the proposal is that boundary entanglement entropy is dual to the generalized entropy of the extremal-area surface: $S_R = S_{\text{gen}}(X)$. The FLM proposal passes some non-trivial consistency checks, but is only supposed to provide the next-to-leading order correction in a $1/N$ expansion. 

A natural extension of this conjecture is presented in \cite{EngWal14}: instead of extremizing the area and solving for the generalized entropy, we find the surface $\chi$ which extremizes the generalized entropy subject to $\partial \chi = \partial R$ and $\chi$ homologous to $R$. The proposal in \cite{EngWal14} is to identify the generalized entropy of $\chi$ with the entanglement entropy of the boundary field theory in the region $R$: $S_R = S_{\text{gen}}(\chi)$. These extremal entropy surfaces are called \textit{quantum extremal surfaces}. While this construction agrees at leading order with the FLM proposal, they differ at higher orders in $N$.  

A classical argument in \cite{EngWal13} shows that assuming the null energy condition, a null surface shot out from a codimension-2 extremal surface acts as a ``barrier'' to other extremal surfaces, in the sense that no continuous 1 parameter family of extremal surfaces can be extended across the barrier.  This result was extended to quantum extremal surfaces in \cite{EngWal14}, but the proof required use of the QFC, in order to show that the null surface shot out from the quantum extremal surface becomes quantum trapped.  So once again, the focussing conjectures bring the quantum generalizations closer in line with the classical result.

\subsection{Generalized Second Law for Quantum Holographic Screens}
\label{sec-screens}

A new classical area law in General Relativity was recently formulated and proven~\cite{BouEng15a,BouEng15b}, assuming the null energy condition. A marginally trapped surface is a compact codimension-2 surface whose classical expansion vanishes in one orthogonal null direction $k^a$ and is strictly negative in the other direction, $l^a$. A future holographic screen is a hypersurface of indefinite signature, foliated by marginally trapped surfaces called ``leaves.'' Subject to certain generic conditions, it was shown that the foliation of a future holographic screen evolves monotonically in the $-l^a$ direction. (E.g., for a black hole formed by collapse, this is the outside or past direction.) This implies further that the area of a future holographic screen increases monotonically along the foliation. A similar area law holds for past holographic screens, defined in terms of marginally anti-trapped surfaces, and abundant in cosmological solutions such as our own universe.

Past or future holographic screens are easily constructed by picking a null foliation of the spacetime. On each codimension-1 null slice one finds the unique codimension-2 surface of maximal area. This surface may lie on the conformal boundary, but if gravity is strong it will lie inside the spacetime. One null expansion vanishes by construction, so the sequence of such surfaces form a holographic screen~\cite{CEB2}. The screen will be future or past if the sign of the other expansion is definite.

The proof of the theorem is elaborate in general but simple in the case of spherical symmetry. It is easy to see that screens that violate the theorem are intersected twice by the same null congruence $N$ with tangent vector $k^a$ on two distinct leaves. The generic condition implies $\theta\neq 0$ between the two leaves. But $\theta=0$ on both leaves, in conflict with the classical focussing theorem, Eq.~(\ref{eq-clfocintro}).

Like Hawking's area theorem for event horizons, the new area law fails when the null energy condition is violated. And like Hawking's theorem, the area theorem for holographic screens can be reformulated as a Generalized Second Law~\cite{BouEngTA}, by replacing area with generalized entropy via Eq.~(\ref{eq-sub}). This is the first covariant statement of a Generalized Second Law that applies to general quasilocal horizons, and the first that applies to expanding cosmologies regardless of the sign of the cosmological constant.

The novel GSL applies to quantum future (or past) holographic screens. These are defined as hypersurfaces foliated by quantum marginally trapped (or antitrapped) surfaces. The latter, in turn, are defined by requiring that the quantum expansion vanishes, $\Theta=0$, in one null direction and is strictly negative, $\Theta<0$, in the other. Like classical holographic screens, these objects are easily constructed in general spacetimes. 

The proof of this novel GSL proceeds exactly as in the classical area theorem, with the assumption of the null energy condition replaced by the QFC. Again, a generic condition implies that $\Theta\neq 0$ on $N$ between two leaves. The definition of the quantum holographic screen requires $\Theta=0$ on both leaves, in contradiction with the QFC~\cite{BouEngTA}.

\acknowledgments It is a pleasure to thank C.~Akers, E.~Bianchi, W.~Donnelly, N.~Engelhardt, B.~Freivogel, M.~Headrick, G.~Horowitz, T.~Jacobson, J.~Koeller, J.~Maldacena, D.~Marolf, and D.~Simmons-Duffin for discussions. The work of RB, ZF, and SL is supported in part by the Berkeley Center for Theoretical Physics, by the National Science Foundation (award numbers 1214644 and 1316783), by fqxi grant RFP3-1323, and by the US Department of Energy under Contract DE-AC02-05CH11231. The work of AW is supported in part by NSF grant PHY-1314311 and the Institute for Advanced Study.

\appendix

\section{Renormalization of the Entropy}
\label{ren}

It is well known that the entanglement entropy $S_\text{out}$ on one side of a sharp boundary $\sigma$ is subject to UV divergences.  Thus in order to define the generalized entropy $S_\text{gen}$, we must invoke a renormalization procedure.  We start by regulating $S_\text{out}$ using a UV cutoff associated with some distance scale $\epsilon$, so that the outside entropy $S_\text{out}^{(\epsilon)}$ now depends on the regulator.  This could be done using e.g. a heat kernel regulator \cite{Vassilevich03, Solodukhin95}, Pauli-Villars \cite{PV49,DLM95}, a brick wall cutoff \cite{tHooft98}, the mutual information \cite{Casini06}, or a variety of other methods.

The leading-order divergence is proportional to the area \cite{Sorkin83, BKLS86, Srednicki93}, but in dimensions $D \ge 4$ there are additional subleading divergences, each proportional to some local geometrical integral on the boundary.  In dimension $D$ there are subleading divergent corrections with weight up to $D$.  Since perturbative quantum gravity is a nonrenormalizable theory, one can also even higher curvature corrections by considering 2-loop or higher diagrams involving gravitons.  Higher curvature corrections can also arise from stringy effects.

When calculating the generalized entropy, each of these divergences is absorbed into a counterterm, i.e., a parameter in the gravitational action $I$ which controls the size of a correction to the gravitational entropy $S_\text{grav}$.  We shall see that the total quantity $S_\text{gen} = S_\text{grav} + S_\text{out}$ is invariant under the RG flow.  Thus these counterterms are important part of the definition of the QFC, although they drop out of the QNEC for the reasons described in section~\ref{sec-diagonal}.

\subsection{The Replica Trick}\label{replica}

The replica trick (reviewed in \cite{CC09}) is a convenient way to calculate the von Neumann entropy $-\text{tr}(\rho \log \rho)$, without the nuisance of taking the logarithm of a matrix.  This trick is based on instead evaluating the Renyi entropy
\begin{equation}
S_n = \frac{1}{1 - n} \ln \mathrm{tr}(\rho^n)
\end{equation}
and analytically continuing to $n = 1$ to obtain $S$.  In cases where the state $\rho$ comes from a Euclidean path integral, there is a beautiful geometrical interpretation of the Renyi entropy in terms of an $n$-sheeted cover $M^{(n)}$ of the manifold, having a conical singularity with total angle $2\pi n$ on the entangling surface $\sigma$.  Assuming that one can analytically continue the effective action $I_\mathrm{eff} = -\ln Z$ from the positive integers to $n = 1$, one then writes
\begin{equation}\label{geometric}
S_\text{replica} = - (1 - n \partial_n) I_\mathrm{eff} \Big|_{n=1}.
\end{equation}
This defines the \textit{geometrical} or \textit{replica} entropy for the state of the quantum fields outside of $\sigma$ \cite{Frolov:1993ym, Susskind:1994sm, Callan:1994py, Holzhey:1994we, Barvinsky:1994jca, Solodukhin:1994yz, Calabrese:2004eu, Calabrese:2005zw, Cardy:2007mb}.  

Inserting the 1-loop effective action $I_\mathrm{eff}$ of a quantum field, we obtain a nonlocal answer for $S_\text{replica}$, as expected.  However, the UV divergences in $I_\mathrm{eff}$ are local, allowing us to compute the corresponding divergences in $S_\text{replica}$.

Note that the identification of $S_\text{replica}$ with the von Neumann entropy $S_\mathrm{out}$ is somewhat formal, due to the fact that the replicated manifold has a delta function of curvature at the conical singularity.\footnote{This curvature is sometimes smoothed out slightly in order to evaluate the entropy \cite{Nelson94}, although this smoothing does not by itself eliminate the UV divergences associated with coupling to the curvature.}  Matter fields can couple to the curvature at the tip, producing ``contact terms'' whose interpretation will be discussed in section \ref{contact}.

Now nothing stops us from inserting a \emph{classical} gravitational action $I[g_{ab}]$ into \eqref{geometric} (treating the n-sheeted cover manifold as a fixed background metric).  In this case everything cancels except for a contribution coming from the conical singularity.  So in this case we obtain an entropy which is local on $\sigma$, which we call the gravitational entropy $S_\text{grav}$.

If $I$ is the Einstein-Hilbert action, Gibbons and Hawking obtained by this method the Bekenstein-Hawking entropy $A/4G\hbar$ \cite{GH76}, thus explaining why this term is included in the generalized entropy.  If on the other hand $I$ is a higher-curvature action, one obtains additional correction terms in $S_\text{grav}$.  For the stationary case, in which $\sigma$ lies on a Killing horizon, the analytic continuation is easy due to the presence of the rotational symmetry about the bifurcation surface, and hence $S_\text{grav}$ is given by the Wald entropy \cite{Wald93, JKM93, IW94, IW95}, obtained by differentiating the Lagrangian with respect to the Riemann tensor and multiplying by the binormal $\epsilon_{\mu \nu}$ twice:
\begin{equation}\label{Wald}
S_\mathrm{Wald} = -\frac{2\pi}{\hbar} \int_\sigma d^2x \sqrt{^2g} \frac{\partial I}{\partial R_{\mu\nu\xi o}} \epsilon_{\mu\nu} \epsilon_{\xi o}.
\end{equation}
However, Wald's Noether charge method for deriving the entropy is subject to ambiguities for non-stationary horizons \cite{JKM93, IW94}, and is therefore unable to determine the coefficients of those terms which vanish on stationary horizons, e.g. terms involving products of extrinsic curvatures.

Thus, for the nonstationary case, there are additional corrections to the Wald entropy which were only recently calculated for higher-curvature gravity actions, which will be discussed in the next section.

\subsection{Nonstationary Entropy}\label{nonstat}

For several years it was unclear how to calculate $S_\mathrm{grav}$ for a nonstationary surface $\sigma$ in a higher-curvature gravity theory, although some information was available using various methods such as field redefinitions and the GSL \cite{JKM93, JKMincrease, FR01}, holography \cite{Solodukhin08, Hung:2011xb}, and the Randall-Sundrum model \cite{MyePou13}.

A major breakthrough came when Lewkowycz and Maldacena \cite{LM13} found a clever way to analytically continue the smoothed out n-sheeted replica trick in the context of calculating the holographic entanglement entropy (cf. section \ref{sec-barrier}) in AdS/CFT.  Their calculation involves performing the replica trick on the conformal boundary of the manifold, while requiring the interior to be a smooth solution to the equations of motion, exploiting the dynamical nature of gravity.  One can then orbifold by the replica group $Z_n$ to find a manifold for which the $n \to 1$ limit can be smoothly taken.  This can be used to derive the Ryu-Takayanagi formula \cite{RyuTak06}\footnote{And presumably also the HRT formula \cite{HubRan07}, if one analytically continues using a complexified manifold.} in the regime where the bulk theory is governed by the Einstein-Hilbert action.

Their calculation was quickly extended to determine the gravitational entropy functional $S_\text{grav}$ for higher-curvature theories.  For quadratic gravity, see \cite{FPS13, Dong13, Camps13}; for Lovelock see \cite{Bhattacharyya:2013jma, Bhattacharyya:2013gra, Bhattacharyya:2014yga}.  In the more general case of f(Riemann) actions, $S_\text{grav}$ is given by the Dong entropy \cite{Dong13}  But some ambiguities remain, related to the ``splitting problem'' \cite{Miao:2014nxa, Miao:2015iba} (these references also make some inroads into the case where the action contains derivatives of the Riemann tensor).

Also, Faulkner, Lewkowycz and Maldacena showed how to include the 1-loop matter in the bulk using \cite{FLM13}; this corresponds to adding a bulk entanglement entropy term, thus replacing $A$ with $S_\text{gen}$ of the extremal surface, as done many times in this article.

Although these formulae were derived for holographic entanglement surfaces, they are consistent with the hypothesis that an entropy can be ascribed to more general surfaces.  For example, the holographic entropy functional also seems to be the correct one to use when defining the GSL for linearized metric perturbations to Killing horizons, in any higher curvature gravity theory \cite{Wall:2015raa} (cf. \cite{Bhattacharjee:2015yaa, Sarkar:2013swa} for some special cases).

\subsection{Example: 3+1 Dimensions}

For example, in $D = 4$ semiclassical gravity coupled to free fields, the one loop corrections to the inverse of Newton's constant $1/G$ are quadratically divergent in $\epsilon$:
\begin{equation}
\Delta_\text{1-loop} \frac{1}{G} = f_G \, \epsilon^{-2},
\end{equation}
where the constant of proportionality $f_G$ depends on the number and type of matter species.  But there are also logarithmic divergences in the three parameters $\alpha, \beta, \gamma$ associated with the quadratic gravity effective action \cite{FPS13}:
\begin{equation}\label{eq:gravdiv1}
I_\mathrm{eff} = \int d^4x\,\sqrt{g} \left [\frac{R}{16\pi G} + \alpha R^2 + \beta (R_{\mu \nu})^2 + \gamma (R_{\mu \nu o \xi})^2 \right] + I_\mathrm{nonlocal}.
\end{equation}where, at one loop,\begin{align}\label{eq:gravdiv2}
\Delta \alpha_\text{1-loop} &= f_\alpha \log(\epsilon) + g_\alpha; \\
\Delta \beta_\text{1-loop} &= f_\beta \log(\epsilon) + g_\beta; \\
\Delta \gamma_\text{1-loop} &= f_\gamma \log(\epsilon) + g_\gamma.
\end{align}
In general, coefficients of power law divergences such as $f_G$ depend on the details of the renormalization scheme.  But the coefficients of the log divergences $f_\alpha,f_\beta,f_\gamma$ are universal, depending only on the field theory \cite{Casini06}.\footnote{In the special case of a CFT these parameters are determined by the central charges $c$ and $a$, which determine the log divergences of the two conformally invariant contributions: Weyl squared $(C_{\mu \nu \alpha \xi})^2$ and the Euler density $R_{\mu \nu \alpha \xi} R_{\pi \rho \sigma \tau} \epsilon^{\mu \nu \pi \rho} \epsilon^{\alpha \xi \sigma \tau}$ respectively.  Although these two terms would be conformal if they were finite, their logarithmic dependence on $\epsilon$ is a conformal anomaly; thus the partition function on curved spacetimes is not scale invariant.}  These log divergences appear because the dimension is even.  (The finite piece of the 1-loop effective action is universal in odd dimensions, but in even dimensions this is true only up to local counterterms such as $g_\alpha,g_\beta,g_\gamma$, since $\epsilon^0$ still counts as a power law!)  The value of the $f$ coefficients for various spins in 4 dimensions are listed in \cite{CD79, BirrellDavies, Vassilevich03}, although there are certain issues with higher spin fields ($3/2$ and $2$) which we will discuss in \ref{spins}.

Dimensional analysis reveals five possible covariant terms in the gravitational entropy density of $\sigma$ due to $\alpha,\,\beta,\,\gamma$:
\begin{equation}
R \,\,\, \qquad R_i^{\phantom{i}i} \qquad R_{ij}^{\phantom{ij}ij} \qquad
K^{\phantom{i}i}_{i\phantom{i}a} K^{\phantom{j}j}_{j\phantom{j}a} \qquad
K_{ij}^{\phantom{ij}a} K^{ij}_{\phantom{ij}a},
\label{eq:action}
\end{equation}
where $R_{\mu \nu \xi o}$ is the 4D Riemann tensor, $R_{\mu \nu}$ the 4D Ricci tensor, $R$ is the 4D Ricci scalar, $K_{ij}^{\phantom{ij}a}$ is the extrinsic curvature; also the indices $i, j$ represent directions parallel to $\sigma$, which are raised and lowered by $q_{ij}$ (the metric restricted to $\sigma$), and $a$ represents an index normal to $\sigma$.  The extrinsic curvature requires an $a$ index because $\sigma$ is a codimension-2 surface, so that there are two remaining dimensions to bend into.

However, because there are only 3 possible terms in the action, only 3 linear combinations of these 5 terms can appear.  The methods of \ref{nonstat} show that the correct entropy functional is \cite{Solodukhin08, MyePou13, FPS13, Dong13, Camps13}
\begin{equation}
S_\text{grav} = \frac{A}{4G\hbar} + \frac{1}{2\pi \hbar} \int_\sigma d^2x \sqrt{^2g} \left[
\alpha R + 
\frac{\beta}{2}
\left(R_i^{\phantom{i}i} \!-\! K^{\phantom{i}ia}_i K^{\phantom{j}j}_{j\phantom{j}a} \right) + 
\gamma \left(R_{ij}^{\phantom{ij}ij} \!-\! K_{ij}^{\phantom{ij}a} K^{ij}_{\phantom{ij}a}\right)
\right]
\end{equation}
Aside from the extrinsic curvature terms (which are ambiguous in the Noether charge formalism), this expression is the same as the Wald entropy \eqref{Wald}.  We may then define the generalized entropy as \cite{DW12}
\begin{equation}\label{Donggen}
S_\text{gen} = \langle S_\text{grav} \rangle + S_\text{out}
\end{equation}
which in this case expands out to:
\begin{eqnarray}
S_\mathrm{gen} = \lim_{\epsilon \to 0} \left[ 
S_\mathrm{out}^{(\epsilon)} + \frac{A}{4G(\epsilon)\hbar}  +
\frac{1}{4\pi \hbar} \int_\sigma d^2x \sqrt{^2g} \, \times \phantom{MMMMM}
\right. \nonumber
\\ 
\left. \left(
\alpha(\epsilon) R + 
\frac{\beta(\epsilon)}{2}
\left(R_i^{\phantom{i}i} - K^{\phantom{i}ia}_i K^{\phantom{j}j}_{j\phantom{j}a} \right) + 
\gamma(\epsilon)\left(R_{ij}^{\phantom{ij}ij} - K_{ij}^{\phantom{ij}a} K^{ij}_{\phantom{ij}a}\right) \right)
\right] ,
\label{eq:gendiv}
\end{eqnarray}
In the limit where $\epsilon$ becomes small, the various dependences on $\epsilon$ should cancel out, so that $S_\mathrm{gen}$ is independent of the choice of cutoff scale.  Since the divergences in $S_\text{out}$ are local and proportional to the other terms in \eqref{eq:gendiv}, it is manifest that there exists a choice of RG flow for the parameters $G,\alpha,\beta,\gamma$ for which $S_\mathrm{gen}$ becomes cutoff independent.

Thus the generalized entropy $S_\text{gen}$ is well-defined even though $S_\text{out}$ is cutoff dependent.  As one shifts the cutoff, the entropy simply moves between the different terms.

\subsection{Interpretation of Contact Terms}\label{contact}

A crucial consistency condition is that the RG flow of the parameters in $S_\text{gen}$ (given by \eqref{eq:gendiv} at one loop in 3+1 dimensions) are in fact the \emph{same} as for the corresponding parameter in the gravitational action \eqref{eq:gravdiv1}.\footnote{For non-universal coefficients, the same regulator must of course be used on both sides.}

Although the renormalization of the replica trick entropy automatically matches the renormalization of the gravitational action \cite{LW95,CL13}, it is not so clear that the replica entropy can always be written as the sum of a horizon piece plus a statistical piece which is literally a von Neumann entropy $-\text{tr}(\rho \log \rho)$, as in \eqref{Donggen}.  Thus if one e.g. imposes a regulator such as a brick wall (or a lattice) in which it is manifest that $S_\text{gen}$ has literally a statistical interpretation, and then compares to the replica trick with e.g. a smoothed out conical singularity, it is not \emph{a priori} obvious that the RG flow of the two definitions of entropy will agree.  The question is whether all terms in the replica entropy can be given a statistical interpretation.

For minimally coupled scalars and spinor fields, several calculations have shown an exact agreement between the geometric and statistical viewpoints \cite{FS94, DLM95, Solodukhin95, dAO95, FFZ95, Kabat95, Winstanley00} (modulo the $K^2$ terms, whose coefficients were unknown in the 1990s but can now be determined and shown to agree by the methods described above).\footnote{But see \cite{KKSY97} for an apparent discrepancy for scalars in odd dimensions, using a Pauli-Villars regulator.}

There is an apparent mismatch \cite{Solodukhin11,DW12} between the replica entropy of non-minimally coupled scalars \cite{Solodukhin95} and gauge fields \cite{Kabat95}, and their statistical entanglement entropy.  Although the renormalization of the replica entropy automatically matches the renormalization of the gravitational action \cite{LW95,CL13}, it is not so clear that this geometrical entropy can always be written as the sum of a horizon piece plus a statistical piece, as in \eqref{Donggen}.  But these concerns can be resolved.

In the case of non-minimally coupled scalar fields whose Lagrangian includes the term $\xi \phi^2 R$, an extra ``contact term'' appears due to coupling to the conical singularity when performing the replica trick.  This extra term is proportional to $\xi \int_\sigma \langle \phi^2 \rangle$.  It contributes to the $S_\text{grav}$ (which in this case is equal to the Wald entropy \eqref{Wald}), and hence appears as an extra term in \eqref{eq:gendiv}.  It also contributes nontrivially to the RG flow of $1/G$, due to the multiplication of $\phi$ at coincident points, restoring consistency \cite{DW12}.

In the case of Maxwell fields, there is also a contact term, which however cannot be explained by the addition of any term of the appropriate dimension to $S_\text{grav}$.\footnote{By analogy to non-minimal scalar, one could try to add a term like $A_i A_j g^{ij}_\perp$, where $A_i$ is the gauge potential and $g^{ij}_\perp$ the inverse normal metric \cite{DW12}, but this is not a gauge-invariant combination.  Nor does it appear in the holographic gravitational entropy \cite{WHHuang14} when derived by the Lewkowycz-Maldacena method \cite{LM13}.}  Instead, the extra term arises from ``edge modes'' in the entanglement entropy due to the fact that boundary gauge symmetries are not gauged, and can be calculated by a path integral over the electric flux through the surface $\sigma$ \cite{DW14} (cf. \cite{KWHuang14}).\footnote{An alternative explanation \cite{Astaneh:2014sma} is that the contact term arises from the total derivative terms in the action.  It has also been suggested \cite{Astaneh:2014wxg, Astaneh:2015tea} that this would resolve a discrepancy between the field theoretical and holographic entropy calculations in 6 dimensions found by \cite{Hung:2011xb}.  However, this interpretation departs from the usual principle that total derivatives may be dropped without consequence, and was argued against in \cite{Huang:2015zua}.}

A similar contact term involving gravitons can presumably be resolved in the same way, but here there are additional conceptual problems which we briefly explore next.

\subsection{The Challenge of Higher Spin Fields}\label{spins}

Defining the entanglement entropy for spin 3/2 and 2 fields is considerably trickier than for spins $\le 1$.

First of all, as a result of their relationship to (super)gravity, they are only consistently defined when expanding around a gravitational background where the traceless part of the Ricci tensor vanishes (Einstein manifolds) \cite{CD79}.\footnote{In addition to these constraints, spin 3/2 fields require the cosmological constant to be negative or zero.  A further issue (unique to fermionic gauge fields) is that when gauge-fixing the spin 3/2 field, one must take into account not only the Faddeev-Popov ghosts but also the Nielsen-Kallosh ghost \cite{Nielsen78, Kallosh78}, in order to obtain the proper number of degrees of freedom \cite{IN85}; see \cite{Will_dissertation} for a correction of the spin 3/2 calculation in Ref.~\cite{Solodukhin11}.}  If the trace is nonzero, i.e., if there is a cosmological constant, the fields acquire an effective mass which must be included as in \cite{CD80}.\footnote{For this reason, among other errors, the $R^2$ coefficient for gravitons or gravitinos found in \cite{CD79, BirrellDavies, Vassilevich03} should not be used.}  This suggests that only some of the coefficients defined above are physically meaningful, namely those of the Weyl squared term and terms involving the Ricci scalar.  As a result, the usual bulk replica trick method for defining the entanglement is suspect, since this is invalid off-shell \cite{CL13}.

A possibly related problem: In the case of gravitons, it is also necessary to decide on a covariant definition of the location of the surface $\sigma$.  Choosing $\sigma$ based on its coordinate location would make the results dependent on the choice of gauge.

In spite of these troublesome issues, some have ploughed ahead and calculated the $f_G$ coefficient for linearized gravitons using a spin-2 gauge-fixed wave operator.  Fursaev and Miele \cite{FM97} found that the partition function of the replica manifold does not reduce in the limit $\beta \to 2\pi$ to the partition function of the smoothed out cone, which usually must agree \cite{Nelson94}.  Solodukhin \cite{Solodukhin11, Solodukhin:2015hma} also calculated $f_G$ for the graviton, but his results appear to conflict with \cite{FM97}.

More ambitiously still, He et al. \cite{He:2014gva} calculated the entanglement entropy for the entire tower of higher-spin fields which appear in string theory, using an alternative definition of the replica trick on orbifold spacetimes where $1/n$ is an integer.  (For spin 3/2 and 2 their calculation agrees with that of \cite{FM97}.)

Notwithstanding the conceptual problems above, it should at least be possible to calculate the graviton entanglement entropy using the Lewkowycz-Maldacena method \cite{LM13}, in which the replica trick is performed on the boundary, and the equations of motion hold everywhere in the bulk.  But this would only allow one to calculate the graviton entropy when $\sigma$ is an extremal surface!  It would not be sufficient for applications such as the QFC in which $\sigma$ may be an arbitrary surface.  Also, in order to derive the result of Faulkner et al. \cite{FLM13} for gravitons, one would need an independent bulk definition of the graviton entanglement entropy.

Thus, further exploration into the nature of entanglement entropy for gravitons and gravitinos seems warranted.

\bibliographystyle{utcaps}
\bibliography{all}

\end{document}